\documentclass[vecphys]{svmult}

\usepackage{makeidx}         
\usepackage{graphicx}        
\usepackage{multicol}        
\usepackage[bottom]{footmisc}
\usepackage{subfigure}

\makeindex             

\begin{document}
\newcommand\subfigsize{4cm}
\setlength\subfigcapskip{10pt}

\title*{Hydrodynamics and Flow}
\author{Tetsufumi Hirano\inst{1}\and
Naomi van der Kolk\inst{2}\and Ante Bilandzic\inst{2}}
\institute{Department of Physics, The University of Tokyo,
Tokyo 113-0033, Japan
\texttt{hirano@phys.s.u-tokyo.ac.jp}\and Nikhef, Kruislaan 409, 1098 SJ Amsterdam, The Netherlands\\ \texttt{kolk@nikhef.nl, anteb@nikhef.nl}}
%
%

\maketitle

\section{Introduction and Disclaimer}
\label{s:mainIntrodution}
The main purpose of the lecture
was to lead students and young post-docs to the frontier of the
hydrodynamic description of relativistic heavy-ion collisions (H.I.C.) in order
for them to understand talks and posters presented in the
Quark Matter 2008 (QM08) conference in Jaipur, India \cite{QM08}. 
So the most recent studies were not
addressed in this lecture as they would be presented during
the QM08 conference itself. Also,
we try to give a very pedagogical lecture here.
For the readers who may want to study relativistic hydrodynamics
and its application to H.I.C. as an advanced course,
we strongly recommend them to consult the references.

This lecture note is divided into three parts. In the first
part we give a brief introduction to relativistic hydrodynamics in
the context of H.I.C. In
the second part we present the formalism and some fundamental
aspects of relativistic ideal and viscous
hydrodynamics. In the third part, we start with some basic checks of
the fundamental observables followed by discussion of
collective flow, in particular \textit{elliptic flow}, which
is one of the most exciting phenomenon in H.I.C. at
relativistic energies. Next we discuss how to formulate the
hydrodynamic model to describe dynamics of H.I.C.
Finally, we conclude the third part of the lecture note by
showing some results from ideal hydrodynamics calculations and by
comparing them with the experimental data.

We use the natural units $c = \hbar = k_{B} = 1$
and the Minkowski metric $g^{\mu\nu} = \mathrm{diag}(1,-1,-1,-1)$ throughout
the lecture note.

\section{Introduction to hydrodynamics in relativistic heavy-ion collisions}
\label{s:introduction}

\begin{figure}
\centering
\includegraphics[width=6cm]{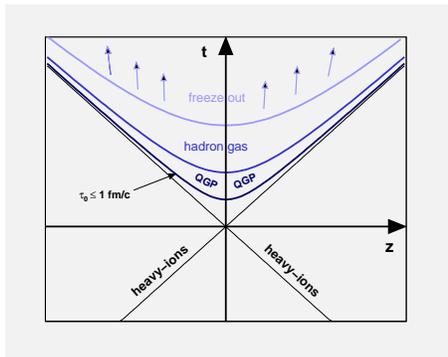}
\caption{A schematic view of dynamics of a heavy ion collision
along the collision axis.}
\label{fig:slide7}
\end{figure}

The excitement raised by the announcement of the discovery of the
``perfect'' liquid at Relativistic Heavy Ion Collider (RHIC) in
Brookhaven National Laboratory (BNL) \cite{BNL}
is based on an agreement
between predictions from \textit{ideal}
hydrodynamic models with the experimental data.
While this agreement was certainly a large boost for various
groups around the world doing research in hydrodynamics
(and even in string theory!), there are
also other reasons why the usage of hydrodynamics is strongly
needed in H.I.C. 
Needless to say, the main goals of the physics of H.I.C.
are to discover the deconfined nuclear matter under equilibrium,
namely the Quark Gluon Plasma (QGP), and to understand its properties
such as equation of state (EoS), temperature and order of phase transition,
transport coefficients and so on.
However, the system produced in H.I.C. dynamically evolves
within time duration of the order of 10-100 fm/$c$.
Hence one has to describe space-time evolution of thermodynamic variables
to fill a large gap 
between the static aspects of QGP properties and dynamical aspects of
H.I.C.
It is the hydrodynamics that plays an
important role in connecting them.
Various stages of H.I.C. are depicted in Fig.~\ref{fig:slide7}.
Two energetic nuclei are coming along light-cone
and collide with each other to create a multi-parton system.
Through secondary collisions the system
may reach thermal equilibrium and the QGP can be formed.
This is a transient state: After further expansion and cooling the system
hadronizes again. Eventually, expansion leads to a free-streaming
stage through freezeout and particle spectra at this moment
are seen by the detector.
Hydrodynamics is applied to matter under local 
equilibrium in the intermediate stage.
Of course, it is non-trivial a priori whether one can always apply hydrodynamics
to the dynamics of H.I.C. Nevertheless it is not a bad idea to dare to apply it
since we eager to understand the matter under equilibrium in
terms of H.I.C.

There is also another good reason to apply hydrodynamics to H.I.C.
A lot of experimental data have been published
so far at various collision energies.
Ideally, one may want to describe these
data from the first principle, \textit{i.e.}, 
quantum chromodynamics (QCD). The QCD
Lagrangean density reads
\begin{equation}
\mathcal{L} = \bar{\psi}_i\left(i\gamma_\mu D^{\mu}_{ij}-m\delta_{ij}\right)\psi_j-\frac14 F_{\mu\nu\alpha}F^{\mu\nu\alpha}\,.
\end{equation}
where $\psi_i$ is a quark field, $D^\mu$ is a covariant derivative, $m$ is a quark mass
and $F^{\mu\nu}_{\alpha}$ is a field strength of gluons. 
However, in spite of its simple-looking Lagrangean,
it is very difficult to make any predictions directly from QCD
in H.I.C.
due to its complexity which mainly arises from 
non-linearity of interactions of gluons, strong coupling,
dynamical many body system and color confinement.
One promising strategy to connect
the first principle with phenomena
is to introduce hydrodynamics as a phenomenological theory.
We call this strategy a bottom-up approach to H.I.C.
An input to
this phenomenological theory comprises the equation of state,
\begin{equation}
P = P(e,n),
\end{equation}
which expresses the pressure $P$ as a function of energy density
$e$ and baryon density $n$.
This can be obtained by exploiting lattice numerical simulations
of QCD\footnote{From lattice calculations, pressure as a function 
of temperature rather than energy density is obtained.
Note also that, due to sign problem,
thermodynamic variables are available only near the region of
vanishing chemical potential.}.
In the case of viscous hydrodynamics
 we need additionally the transport coefficients such as shear
viscosity $\eta$, bulk viscosity $\zeta$, heat conductivity
$\lambda$, etc.\footnote{In principle, 
the information about these quantities
can be obtained also from the lattice QCD simulations
although it is much harder than the EoS.}

Once hydrodynamics turns out to work
quite well in description
of dynamics, one can utilize its outputs
such as local temperature or energy density
for other observables.
In the current formalism of jet quenching,
one needs an information of parton density or energy density
along a trajectory of an energetic parton \cite{Gyulassy:2003mc,Kovner:2003zj}.  
If one assumes $J/\psi$ melts away above some temperature \cite{Matsui:1986dk},
one needs local temperature at the position of $J/\psi$.
In the case of electromagnetic probes \cite{Peitzmann:2001mz,d'Enterria:2005vz},
one convolutes emission rate (the number of produced particles
per unit space-time volume at temperature $T$)
of thermal photons and dileptons
over the space-time volume under equilibrium.
Hydrodynamics provides us with the information of the bulk matter.
Therefore we can say that, in the context of H.I.C., hydrodynamics is the heart of
the dynamical modeling: It not only describes
expansion and collective flow of matter but also provides important informations
in the intermediate stage for other phenomena.

\section{Formalism of the relativistic ideal/viscous hydrodynamics}
\label{s:formalism}

The second part of the lecture note is more formal with many
equations, but we try as much as possible to provide the intuitive
picture behind the equations. The following references
might be very helpful to complement this section 
\cite{Eckart:1940te,Cattaneo,LL,namiki,namiki2,IMuller,Israel:1976tn,Israel:1979wp,Bjorken:1982qr,Hosoya:1983xm,Danielewicz:1984ww,Csernai,Blaizot:1990zd,Maartens:1996vi,Rischke:1998fq,Muller:1999in,Ollitrault:2007du}.

\subsection{The basic equations}
\label{ss:basic}

The basic hydrodynamical equations
are energy-momentum conservation
\begin{equation}
\partial_\mu T^{\mu\nu} = 0\,,\label{conservationT}
\end{equation}
where $T^{\mu\nu}$ is the energy-momentum tensor, and the current conservation
\begin{equation}
\partial_\mu N_i^\mu = 0\,,\label{conservationN}
\end{equation}
where $N_i^\mu$ is the $i$-th conserved current. 
In H.I.C., there are some conserved charges such as baryon number, strangeness,
electric charges and so on.
We mainly assume the net baryon current $N_B^\mu$ as an example of $N_i^\mu$
in the following. In the
first step we decompose the energy-momentum tensor and the conserved current
as follows:
\begin{eqnarray}
T^{\mu\nu} &=& eu^\mu u^\nu - P\Delta^{\mu\nu} + W^\mu u^\nu + W^\nu u^\mu + \pi^{\mu\nu}\,,\label{decompositionT}\\
N_i^\mu    &=& n_i u^\mu + V_i^\mu\,.\label{decompositionN}
\end{eqnarray}
All the terms in the above expansion will be discussed one by one later. Now
we indicate that $u^\mu$ is the time-like, normalized
four-vector
\begin{equation}
u_\mu u^\mu = 1\,,
\end{equation}
while the tensor $\Delta^{\mu\nu}$ is defined in the following
way,
\begin{equation}
\Delta^{\mu\nu} = g^{\mu\nu} - u^\mu u^\nu\,,
\label{def:Delta}
\end{equation}
where $g^{\mu\nu}$ is the Minkowski metric.
We refer to $u^\mu$ and $\Delta^{\mu\nu}$ as the ``projection''
vector and tensor operators, respectively. In particular, $u^\mu$
is the local flow four-velocity, but a more precise meaning will
be given later. $u^\mu$ is perpendicular to $\Delta^{\mu\nu}$, as
can easily be seen from the definition of $\Delta^{\mu\nu}$ given
in Eq.~(\ref{def:Delta}) and from the fact that $u^\mu$ is normalized,
\begin{equation}
u_\mu \Delta^{\mu\nu} = u_\mu(g^{\mu\nu}-u^\mu u^\nu) = u^\nu - 1\cdot u^\nu = 0\,.
\end{equation}
Next we define the local rest frame (LRF) as the frame in which
$u^\mu$ has only the time-like component non-vanishing and in
which $\Delta^{\mu\nu}$ has only the space-like components
non-vanishing, \textit{i.e.},
\begin{eqnarray}
u^{\mu}_{\mathrm{LRF}} &=& (1,0,0,0)\,,\\
\Delta^{\mu\nu}_{\mathrm{LRF}} &=& \mathrm{diag}(0,-1,-1,-1)\,.
\end{eqnarray}
As is easily understood from the above equations,
one can say that $u^\mu (\Delta^{\mu\nu})$ picks
up the time-(space-)like component(s) when acting on some Lorentz
vector/tensor.

We now discuss the physical meaning of each term in the expansion
of the energy-momentum tensor (\ref{decompositionT}) and the conserved current
(\ref{decompositionN}).

\subsubsection{Decomposition of $T^{\mu\nu}$}
The new quantities which appear on the RHS in the decomposition
(\ref{decompositionT}) are defined in
the following way:
\begin{eqnarray}
e=u_\mu T^{\mu\nu}u_\nu & &\mathrm{(energy\ density)} \,,\\
P=P_s+\Pi  = -\frac13\Delta_{\mu\nu}T^{\mu\nu}& &\mathrm{(hydrostatic+bulk\ pressure)} \,,\\
W^\mu=\Delta^{\mu}_{\ \alpha}T^{\alpha\beta}u_\beta& &\mathrm{(energy\ (or\ heat)\ current)} \,,\\
\pi^{\mu\nu}=\left<T^{\mu\nu}\right>& &\mathrm{(shear\ stress\ tensor)} \,.
\end{eqnarray}
Each term corresponds to projection of the energy momentum tensor by one or two
projection operator(s), $u^{\mu}$ and $\Delta^{\mu\nu}$.
The first two equalities imply that the energy density $e$ can be
obtained from the time-like components of the energy-momentum
tensor, while the pressure $P$ is obtained from the space-like
components. Contracting the energy-momentum tensor simultaneously
with $u^\mu$ and $\Delta^{\mu\nu}$ gives the energy (heat) current
$W^\mu$. Finally, the angular brackets in the definition of the shear
stress tensor $\pi^{\mu\nu}$ stand for the following operation,
\begin{equation}
\left<A^{\mu\nu}\right> = \left[\frac12(\Delta^{\mu\ }_{\ \alpha}\Delta^{\nu\ }_{\ \beta}+\Delta^{\mu\ }_{\ \beta}\Delta^{\nu\ }_{\ \alpha})-\frac13\Delta^{\mu\nu}\Delta_{\alpha\beta}\right]A^{\alpha\beta}\,.
\end{equation}
This means that $\left<A^{\mu\nu}\right>$ is a symmetric and
traceless tensor which is transverse to $u^\mu$ and $u^\nu$.
More concretely, 
one can first decompose the energy momentum tensor by two projection tensors symmetrically
\begin{eqnarray}
\tilde{\pi}^{\mu \nu} & = & 
\frac12(\Delta^{\mu\ }_{\ \alpha} T^{\alpha \beta} \Delta_{\beta}^{\ \ \nu}
+\Delta^{\nu\ }_{\ \alpha}T^{\alpha \beta}\Delta^{\ \ \mu}_{\beta})
\end{eqnarray}
and then decompose it once more into the shear stress tensor (traceless)
and the pressure (non-traceless)
\begin{eqnarray}
\tilde{\pi}^{\mu \nu} & = & \pi^{\mu \nu} - P \Delta^{\mu \nu}.
\end{eqnarray}

\subsubsection{Decomposition of $N^{\mu}$}
In the decomposition (\ref{decompositionN}) we
have introduced the following quantities,
\begin{eqnarray}
n_i=u_\mu N^{\mu}_i & &\mathrm{(charge\ density)}\,,\\
V_i^\mu = \Delta^{\mu\ }_{\ \nu} N_i^\nu && \mathrm{(charge\ current)}\,.
\end{eqnarray}
The physical meaning of $n_i$ and $V_i^\mu$ is self-evident
from the properties of projection operators.

\vspace{0.5cm}
\noindent\textbf{QUESTION 1:} \textit{Count the number of unknowns
in the above decompositions and confirm
that it is $10(T^{\mu\nu})+4k(N_i^\mu)$.
Here $k$ is the number of independent currents}\footnote{
If you consider $u^\mu$ as independent variables,
you need additional constraint for them since these are redundant ones.
If you also consider $P_s$ as an independent
variable, you need the equation of state $P_s=P_s(e,n)$.}.
\vspace{0.5cm}

\noindent The various terms appearing in the decompositions
(\ref{decompositionT}) and (\ref{decompositionN}) can be grouped
into two distinctive parts, which we call ideal and dissipative
part. In particular, for the energy momentum tensor we have,
\begin{eqnarray}
T^{\mu\nu} &=& T_0^{\mu\nu} + \delta T^{\mu\nu}\,,\\
T_0^{\mu\nu} &=& eu^\mu u^\nu - P_s\Delta^{\mu\nu}\,,\label{idealT}\\
\delta T^{\mu\nu} &=& -\Pi\Delta^{\mu\nu}+W^\mu u^\nu+W^\nu u^\mu + \pi^{\mu\nu}\,,\label{dissipativeT}
\end{eqnarray}
while for one charge current we have,
\begin{eqnarray}
N^\mu &=& N_0^\mu + \delta N^\mu\,,\\
N_0^\mu &=& nu^\mu\,,\label{idealN}\\
\delta N^\mu &=& V^\mu\,.\label{dissipativeN}
\end{eqnarray}
In the above relations $T_0^{\mu\nu}(N_0^\mu)$ denote the ideal
part,  while the $\delta T^{\mu\nu}(\delta N^\mu)$ denote the
dissipative part of the $T^{\mu\nu}(N^\mu)$.

\subsection{The meaning of $u^\mu$}
\label{ss:meaning}

\begin{figure}
\centering
\includegraphics[height=4cm]{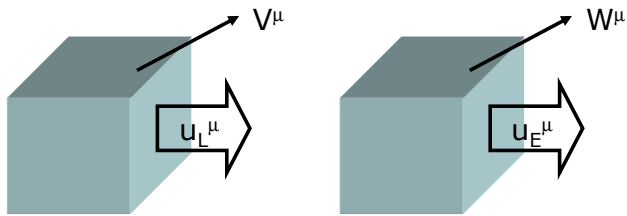}
\caption{A sketch of Landau and Eckart definitions of flow.
Two boxes are fluid elements. There is a ``leak" current
$W^\mu$ or $V^\mu$
according to the definition of flow.}
\label{fig:slide18}
\end{figure}

As we have already mentioned in Sec.~\ref{ss:basic}, $u^\mu$ is the four-velocity of ``flow''.
Now we would like to clarify what kind of flow we have in mind in this description. In literature two definitions of flow can be found,\\
\begin{enumerate}
\item flow of energy (Landau) \cite{LL}:
\begin{equation}
\label{eq:landau}
u_L^\mu=\frac{T^{\mu\ }_{\ \,\nu}u_L^{\nu}}{\sqrt{u_L^\alpha T_{\alpha\ }^{\ \beta}T_{\beta\gamma}u_L^{\gamma}}}=\frac{1}{e}T^{\mu\ }_{\ \,\nu}u_L^{\nu}\,,
\end{equation}
\item flow of conserved charge (Eckart) \cite{Eckart:1940te}:
\begin{equation}
u_E^\mu=\frac{N^\mu}{\sqrt{N_\nu N^\nu}}\,,
\end{equation}
\end{enumerate}
\noindent (see Fig.~\ref{fig:slide18})\footnote{Other definitions
can be made. The situation here is quite similar to
the gauge fixing condition in gauge theories
to eliminate the redundant variables.
An essential point is to choose some ``gauge" for later convenience.
}.
In the first definition, $u^\mu_{L}$ also appears in the RHS of Eq.~(\ref{eq:landau}).
So it should be understood as an equation with respect to $u^\mu_{L}$.
One may solve an eigenvalue problem for a given energy-momentum tensor $T^{\mu\ }_{\ \,\nu}$.
$u^\mu_{L}$ is a normalized time-like eigenvector and the corresponding positive eigenvalue
is energy density $e$.
If the dissipative currents are small enough,
one can show the following relation between these two definitions of flow
\begin{eqnarray}
\label{eq:L-E}
u^\mu_{L}  \approx  u^\mu_{E} + \frac{W^\mu}{e+P_s}, 
\enskip u^\mu_{E} \approx u^\mu_{L}+\frac{V^\mu}{n}.
\end{eqnarray}
Equation (\ref{eq:L-E}) can be shown by assuming that both two definitions of flow can be
connected by infinitesimal proper Lorentz transformation
\begin{eqnarray}
u^\mu_{E} & = & a^{\mu}_{\enskip \nu} u^\nu_{L} \\
& \approx & (\delta^{\mu}_{\enskip \nu}+ \epsilon^{\mu}_{\enskip \nu}) u^\nu_{L}
\end{eqnarray}
where $\epsilon^{\mu}_{\enskip \nu}$ infinitesimal anti-symmetric tensor
and neglecting the higher orders of dissipative currents.
Obviously, $W^\mu = 0$ ($V^\mu = 0$) in the Landau (Eckart) frame.
In the case of vanishing dissipative currents, both definitions represent
a common flow. In other words, flow is uniquely determined 
in the case of ideal hydrodynamics.
We should emphasize that
Landau definition is more relevant in the context of H.I.C. at ultrarelativistic energies since we expect 
a small baryon number is deposited near the midrapidity region.

\subsection{Entropy}

We start this subsection by briefly discussing the entropy
conservation in ``ideal hydrodynamics". By ``ideal hydrodynamics"
we mean the case when entropy is not produced during the evolution\footnote{Note that,
 if discontinuities exist in the solution, entropy is
produced even in ideal hydrodynamics.}.
Neglecting the dissipative parts, the energy-momentum conservation
(\ref{conservationT}) and the current conservation
(\ref{conservationN}) reduce to
\begin{eqnarray}
\partial_\mu T_0^{\mu\nu} &=& 0\,,\label{conservationIdealT}\\
\partial_\mu N_0^\mu &=& 0\,,\label{conservationIdealN}
\end{eqnarray}
where $T_0^{\mu\nu}$ and $N_0^\mu$ are the ideal parts introduced
in Eqs.~(\ref{idealT}) and (\ref{idealN}). Equations
(\ref{conservationIdealT}) and (\ref{conservationIdealN}) are the
basic equations of ``ideal hydrodynamics".

By contracting Eq.~(\ref{conservationIdealT}) with $u_\nu$ it follows,
\begin{eqnarray}
0&=&u_\nu \partial_\mu T_0^{\mu\nu}\nonumber\\
 &=& \ldots \nonumber\\
 &=& T(u^\mu \partial_\mu s + s\partial_\mu u^\mu) +\mu(u^\mu \partial_\mu n + n\partial_\mu u^\mu)\,.\label{entropyConservation}
\end{eqnarray}
We have introduced here the temperature $T$, entropy density $s$
and chemical potential $\mu$
through the first law of thermodynamics $de = Tds + \mu dn$.
Here it is assumed that thermalization is maintained locally.
The second term on the RHS in
Eq.~(\ref{entropyConservation}) vanishes due to
Eq.~(\ref{conservationIdealN}). If we now introduce the entropy
current as
\begin{equation}
S^\mu = su^\mu\,,
\end{equation}
it follows from Eq.~(\ref{entropyConservation}) that
\begin{equation}
\partial_\mu S^\mu = \partial_\mu(su^\mu) = u^\mu \partial_\mu s + s\partial_\mu u^\mu = 0\,.
\end{equation}
Hence the entropy is conserved in ideal hydrodynamics.

\vspace{0.5cm}
\noindent\textbf{QUESTION 2:} \textit{Go through all steps in the above derivations}.
\vspace{0.5cm}

\noindent Now we go back to viscous hydrodynamics. 
Hereafter we consider only the Landau frame and omit the subscript
$L$. For simplicity, we further assume that there is no charge in
the system although in the realistic case a small amount
of charge might exist in the system. What we are constructing here is the so-called first order theory of viscous
hydrodynamics. The main assumption is that the non-equilibrium
entropy current vector $S^\mu$ has linear dissipative term(s)
constructed from $V^\mu$, $\Pi$ and $\pi^{\mu\nu}$ and can be
written as
\begin{equation}
S^\mu = su^\mu + \alpha V^\mu\,.\label{entropyCurrent}
\end{equation}
The first term on the RHS is the ideal part and the second term is
the correction due to the dissipative part. 
It is impossible to construct
a term which would
form a Lorentz vector from $\pi^{\mu\nu}$ on the RHS
in the above equation because $\pi^{\mu\nu}$
is perpendicular to $u^\mu$ by definition\footnote{
Also remember $W^\mu = 0$ in the Landau definition.
One may
think that $\Pi u^\mu$ is a possible candidate in the entropy current $S^\mu$.
However, the second law of thermodynamics is not ensured in this case. See
also discussion in Ref.~\cite{Maartens:1996vi}.}.
Since we have also assumed that there
is no charge in the system, \textit{i.e.}, $N^\mu=0$, it follows that $\alpha
V^{\mu}$ vanishes.

We now calculate the product of the temperature $T$ and the divergence of
the entropy current (\ref{entropyCurrent}). It follows,
\begin{eqnarray}
T\partial_\mu S^\mu &=& T(u^\mu\partial_\mu s + s\partial_\mu u^\mu)\nonumber\\
&=&u_\nu\partial_\mu T_0^{\mu\nu}\nonumber\\
&=&-u_\nu\partial_\mu \delta T^{\mu\nu}\nonumber\\
&=&\ldots\nonumber\\
&=&\pi_{\mu\nu}\left<\nabla^\mu u^\nu\right>-\Pi\partial_\mu u^\mu\,.
\label{TdivS}
\end{eqnarray}
where $\nabla^\mu=\Delta^{\mu\nu}\partial_\nu$.
In transferring from the second to third line in the above calculation
we have used the energy-momentum conservation, $\partial_\mu
T^{\mu\nu}=0$. It is very important to note that due to the
assumption that there is no charge in the system we could neglect
the dissipative part of entropy current (\ref{entropyCurrent}),
but the dissipative part of energy-momentum tensor
(\ref{dissipativeT}) does not vanish. The non-vanishing
dissipative part of energy-momentum tensor gives a contribution
which yields a difference between the equations characterizing the
first order theory of viscous hydrodynamics and the
equations of ideal hydrodynamics derived before.

\vspace{0.5cm}
\noindent\textbf{QUESTION 3:} \textit{Check the above calculation}.
\vspace{0.5cm}

\noindent In order to solve the hydrodynamic equations we must
first define the dissipative current. We introduce the following
two phenomenological definitions, so-called \textit{constitutive
equations}, for the shear stress tensor $\pi^{\mu\nu}$ and the
bulk pressure $\Pi$,
\begin{eqnarray}
\pi^{\mu\nu} &=& 2\eta\left<\nabla^{\mu}u^\nu\right>\,,\label{sst}\\
\Pi &=& -\zeta \partial_\mu u^\mu = -\zeta \nabla_\mu u^\mu\,.\label{bp}
\end{eqnarray}
In Table~\ref{tab:1} we outline the new variables and terminology used in the above equations.
Notice that, within our approximation $N^\mu = 0$, there is no vector component
of thermodynamics force.
\begin{table}
\caption{New variables and terminology.}
\label{tab:1}
\begin{center}
\begin{tabular}{|c|c|c|}
\hline
Thermodynamic force  & Transport coefficient & Current\\
\hline\hline
$X^{\mu\nu}=\left<\nabla^{\mu}u^\nu\right>$ & $\eta$ & $\pi_{\mu\nu}$ \\
tensor & shear viscosity &\\\hline
$X=-\partial_\mu u^\mu$ & $\zeta$ & $\Pi$ \\
scalar & bulk viscosity &\\\hline
\end{tabular}
\end{center}
\end{table}

\noindent After inserting the definitions (\ref{sst}) and (\ref{bp}) in the last line of (\ref{TdivS}),
we arrive at, for positive transport coefficients,
\begin{eqnarray}
T\partial_\mu S^\mu & = & \frac{\pi_{\mu\nu}\pi^{\mu\nu}}{2\eta}+\frac{\Pi^2}{\zeta}\nonumber \\
                    & = & 2\eta \left<\nabla^{\mu}u^\nu\right>^2 + \zeta
                    \left(-\partial_\mu u^\mu \right)^2 \geq 0\,.
\end{eqnarray}
This ensures the second law of thermodynamics
\begin{equation}
\partial_\mu S^\mu \geq 0\,.
\end{equation}
In the case of viscous hydrodynamics, entropy is not decreasing.

\subsection{The equations of motion}
In order to derive the equations of motion, we use again energy-momentum conservation (\ref{conservationT}). After
contracting Eq.~(\ref{conservationT}) with $u_\nu$ we have
\begin{equation}
u_\nu\partial_\mu T^{\mu\nu}=0\,,
\end{equation}
from which one can obtain the first equation of motion,
\begin{equation}
\dot{e}=-(e+P_s+\Pi)\theta+\pi_{\mu\nu}\left<\nabla^{\mu}u^{\nu}\right>\,.
\label{eom1}
\end{equation}
On the other hand, after contracting Eq.~(\ref{conservationT}) with $\Delta_{\mu\alpha}$ it follows,
\begin{equation}
\Delta_{\mu\alpha}\partial_\beta T^{\alpha\beta}=0\,,
\end{equation}
from which one can obtain the second equation of motion,
\begin{equation}
(e+P_s+\Pi)\dot{u}^\mu=\nabla^\mu(P_s+\Pi)-\Delta^{\mu\alpha}\nabla^\beta\pi_{\alpha\beta}+\pi^{\mu\alpha}\dot{u}_\alpha\,.
\label{eom2}
\end{equation}
This is exactly the relativistic extension of the Navier-Stokes equation. In writing the above equations we have introduced,
\begin{eqnarray}
\theta &=&\partial_\mu u^\mu\ \qquad \mathrm{expansion\ scalar\ (divergence\ of\ flow)},\\
\mathrm{``dot"} &=& D = u_\mu \partial^\mu\ \qquad\mathrm{substantial\ time\ derivative}\,.
\end{eqnarray}

\vspace{0.5cm}
\noindent\textbf{QUESTION 4:} \textit{Starting from the energy-momentum conservation} (\ref{conservationT}) \textit{derive equations} (\ref{eom1}) \textit{and} (\ref{eom2}).
\vspace{0.5cm}

\noindent To get some intuitive interpretation of the first
equation of motion, we  insert expressions
(\ref{sst}) and (\ref{bp}) for the shear stress tensor and bulk
pressure into Eq.~(\ref{eom1}),
\begin{eqnarray}
\dot{e}&=&-e\theta-P_s\theta+\frac{\Pi^2}{\zeta}+\frac{\pi_{\mu\nu}\pi^{\mu\nu}}{2\eta}\nonumber\\
&=&-e\theta-P_s\theta+\zeta(-\theta)^2+2\eta\left<\nabla^\mu u^\nu\right>^2\,.
\label{dote}
\end{eqnarray}
The above equation determines the time evolution of energy density
$e$ in the co-moving system.
The first term on the RHS describes dilution/compression of energy density
due to the change of volume,
because $\theta$ can be expressed in terms of volume of a fluid element $V$
as
\begin{equation}
\theta \approx \frac{\dot{V}}{V}\,.
\end{equation}
In ideal hydrodynamics, this relation holds exactly.
If the system expands ($\theta>0$), the energy density is diluted.
So the effect of expansion appears as negative source term
$-e \theta$ in Eq.~(\ref{dote}).
 If we move
along with a fluid element, the internal energy in the fluid element is not conserved
due to the work done by pressure, which is described by the second
term on the RHS in (\ref{dote}). Finally, the last two
positive definite terms in
(\ref{dote}) represent the production of entropy which heats up the
system.

Now we comment on the second equation of motion (\ref{eom2}).
But before doing that, we recall the non-relativistic
Navier-Stokes equation,
\begin{equation}
D\vec{v}=-\frac{1}{\rho}\vec{\nabla}P_s+\frac{\eta}{\rho}\vec{\nabla}^2 \vec{v}\,.\label{nrnse}
\end{equation}
Here $\rho$ is the mass density, 
$\eta$ is shear viscosity and
$D=\frac{\partial}{\partial t} + \vec{v}\cdot\vec{\nabla}$ is 
the non-relativistic version of substantial time derivative.
The above version of the non-relativistic Navier-Stokes equation
applies to the case of incompressible fluids such that
$\vec{\nabla}\cdot \vec{v}=0$ is valid. On the LHS we have the time
derivative of velocity, which is nothing but acceleration. The
first term on the RHS is the source of the flow and it is solely
due to the pressure gradient $\vec{\nabla}P_s$, while the second
term represents the diffusion of the flow. The final flow velocity
comes from the interplay between these two terms: The first term
generates the flow, while the second term dilutes it.
The ratio $\eta/\rho$ is called
kinetic viscosity and plays a role of diffusion constant in
the Navier-Stokes equation (\ref{nrnse}).
The diffusion term in Eq.~(\ref{nrnse}) requires more detailed treatment.
For an illustrative purpose, consider first the heat equation in ($N$+1)-dimensional space-time
\begin{equation}
\frac{\partial T(t,\{x_i\})}{\partial t} = \kappa \sum_i^N \frac{\partial^2}{\partial x_i^2} T(t,\{x_i\})\,,
\label{heat}
\end{equation}
where $T$ is temperature and constant $\kappa$ is heat
conductivity in some unit. One can discretize the heat
equation (\ref{heat}) in $(2+1)$-dimensional space-time:
\begin{eqnarray}
\label{eq:disc_heat}
T_{i,j}^{n+1} & = & T_{i,j}^{n}+\frac{4\kappa\Delta t}{(\Delta x)^2}\left[\frac{T_{i-1,j}^{n}+T_{i,j-1}^{n}+T_{i+1,j}^{n}+T_{i,j+1}^{n}}{4}-T_{i,j}^{n}\right] \nonumber \\
& = & T_{i,j}^{n}+\frac{4\kappa\Delta t}{(\Delta x)^2}\left(\bar{T}_{i,j}^{n}-T_{i,j}^{n}\right)\,,
\end{eqnarray}
where $i$ and $j$ are indices of the site and $n$ is the time step.
The first term in the brackets in Eq.~(\ref{eq:disc_heat}), $\bar{T}_{i,j}$, indicates
an average of temperature around the cell under consideration. If 
temperature at ($i$, $j$) is smaller (larger) than the averaged one 
$\bar{T}_{i,j}>T_{i,j}$ ($\bar{T}_{i,j}<T_{i,j}$),
the second term in Eq.~(\ref{eq:disc_heat}) becomes positive (negative) and, consequently,
temperature increases (decreases) in the next time step.
Repeating this procedure, temperature becomes flat even if starting from
a bumpy initial condition.
Thus, generally speaking, the second derivative with respect to coordinates
describes averaging/smoothening/diffusion of given distributions
and a coefficient in front of it describes how quick the distribution diffuses.
Now going back to the Navier-Stokes equation (\ref{nrnse}),
it is obvious from the above discussion that the second term describes diffusion of flow
and that kinetic viscosity $\eta/\rho$ plays a role of a diffusion constant.
The relativistic version of Navier-Stokes equation (\ref{eom2})
has a similar form to Eq.~(\ref{nrnse}) if one plugs in constitutive equations
(\ref{sst}) and (\ref{bp}) and assumes the fluid is incompressible, $\theta=0$.

\subsubsection{Bjorken's equation in the $1^{\mathrm{st}}$ order theory}
Now we rewrite again the first equation of motion by making use of Bjorken's ansatz~\cite{Bjorken:1982qr}
\begin{eqnarray}
u^\mu_{\mathrm{Bj}}&=& \frac{\tilde{x}^\mu}{\tau} = \frac{t}{\tau}\left(1,0,0,\frac{z}{t}\right)\,.
\end{eqnarray}
where $\tau=\sqrt{t^2-z^2}$ is the proper time.
This is a boost invariant Bjorken's
solution which is also called 1-dimensional Hubble flow since velocity in the $z$ direction,
$v_z$, is proportional to $z$, which is an analogy to 
three dimensional Hubble flow of the universe. 
After inserting this
solution into the constitutive equations (\ref{sst}) and (\ref{bp}) 
\begin{eqnarray}
\pi^{\mu\nu} & = & \frac{2\eta}{\tau} \left(\tilde{\Delta}^{\mu\nu} -\frac{1}{3}\Delta^{\mu \nu}\right)\,,\\
\tilde{\Delta}^{\mu\nu} & = & \tilde{g}^{\mu\nu}-u_{\mathrm{Bj}}^\mu u_{\mathrm{Bj}}^\nu\,,\quad
\tilde{g}^{\mu\nu} \enskip = \enskip \mathrm{diag}(1,0,0,-1)\,,\\
\Pi& = & -\frac{\zeta}{\tau}\,,
\end{eqnarray}
we arrive at
the following equation of motion 
\begin{equation}
\frac{de}{d\tau}=-\frac{e+P_s}{\tau}\left(1-\frac{4}{3\tau T}\frac{\eta}{s}-\frac{1}{\tau T}\frac{\zeta}{s}\right)\,.
\label{be1}
\end{equation}
This equation determines the time evolution of energy density in the $1^{\mathrm{st}}$ order theory
in 1-dimensional expansion.

\vspace{0.5cm}
\noindent\textbf{QUESTION 5:} \textit{Derive equation} (\ref{be1}).
\vspace{0.5cm}

\noindent On the RHS of (\ref{be1}) we have three terms in the bracket. If we
neglect the last two terms this equation reduces to the famous
Bjorken equation \cite{Bjorken:1982qr} which states that in ideal hydrodynamics the
energy density evolution is determined by the sum of energy
density $e$ and the hydrostatic pressure $P_s$, divided by the
proper time $\tau$. The last two terms on the RHS in (\ref{be1})
represent the viscous correction to ideal hydrodynamics. The first one
is the viscous correction originating from the shear viscosity in
compressible fluids, while the second one comes from the bulk
viscosity. We remark that both terms are proportional to $1/\tau$
which is due to the fact that the expansion scalar $\theta$ in the
Bjorken scaling solution can be written as
\begin{equation}
\theta=\frac{1}{\tau}\,.
\end{equation}
Two transport coefficients in the viscous correction, $\eta/s$ and
$\zeta/s$, turn out to be very important.
They are the dimensionless quantities in
natural units and reflect the intrinsic properties of
the fluids\footnote{We stress that in the context of
H.I.C. the statement which is often used, ``viscosity is small",
is not precise. From the equations we have derived we see that the correct
statement should be ``viscous coefficients are small in comparison
with entropy density".}.

Recently progress has been made in obtaining the values of the
transport coefficients from microscopic theories. Here we
summarize the most important results and conclusions,
\begin{itemize}
\item $\eta/s=1/4\pi$ and $\zeta/s=0$ are obtained from $\mathcal{N}=4$ SUSY Yang-Mills theory~\cite{Kovtun:2004de}.
The latter one is automatically obtained from the conformal nature of the theory;
\item $\eta/s=\mathcal{O}(0.1-1)$ for gluonic matter
is obtained from the lattice calculations of pure SU(3) gauge theory \cite{Nakamura:2004sy};
\item Bulk viscosity has a prominent peak around $T_c$
resulting from trace anomaly of QCD \cite{Kharzeev:2007wb,Karsch:2007jc} (see also
a phenomenological approach in Ref.~\cite{Mizutani:1987wb}).
\end{itemize}
\subsection{The $2^{\mathrm{nd}}$ order theory and its application to
Bjorken's equation}

There is an important issue in the first order theory
which is the violation of causality. We
can trace back the origin of the violation of causality to our
phenomenological definitions (\ref{sst}) and (\ref{bp}) for the
shear stress tensor and the bulk pressure, respectively, and to
the fact that the Navier-Stokes equation is a parabolic equation,
namely, the time derivative is
of first order, while the space derivative is of second
order. The same arguments hold also for the violation of causality
in relativistic hydrodynamics: It is known that,
under linear perturbations on the moving background equilibrium state, 
the solutions are unstable and acausal \cite{HiscockLindblom} 
(for a more detailed discussion, see also 
a recent study in Ref.~\cite{Denicol:2008ha}).
For an illustrative purpose, we continue this discussion by
analyzing the heat equation
as an example of the parabolic equation
in three dimensional space\footnote{Again, we choose some unit to simplify the following
equations.},
\begin{equation}
\frac{\partial T}{\partial t}=\kappa\sum_i^{3}\frac{\partial^2}{\partial x_i^2} T\,.
\label{heat2}
\end{equation}
The heat equation can be easily derived by combining the balance equation,
\begin{equation}
\frac{\partial T}{\partial t}= -\frac{\partial q^i}{\partial x^i}\,,
\end{equation}
together with the constitutive equation,
\begin{equation}
q^{i}=-\kappa\frac{\partial T}{\partial x_i}\qquad\mathrm{Fourier's\ law}.
\label{fourier}
\end{equation}
In the above equations $T$ is the temperature, $q^i$ is the
heat current and $\kappa$ is the heat conductivity. The
above constitutive equation is purely phenomenological. Although
we are here considering the non-relativistic equations, the
general arguments and conclusions we write down are valid in
the relativistic case as well. The heat equation (\ref{heat2})
violates causality.
It can be easily confirmed that the Green's function of the heat
equation (\ref{heat2}), sometimes called heat kernel,
is Gaussian
\begin{equation}
G(x^i, t ; x^i_0, t_0) = \frac{1}{[4\pi \kappa (t-t_0)]^{\frac{3}{2}}}
\exp\left[-\frac{(x^i-x^i_0)^2}{4\kappa (t-t_0)} \right]
\label{heatkernel}
\end{equation}
 and the ``long tail'' of this
Gauss function causes the violation of causality in the heat
equation. This issue was heuristically resolved by Cattaneo in 1948 \cite{Cattaneo} after
 an additional term on the LHS of the
constitutive equation (\ref{fourier}) was introduced ``by hand'',
\begin{equation}
\tau_r \frac{\partial q^{i}}{\partial t}+q^{i}=-\kappa\frac{\partial T}{\partial x_{i}}\,.
\end{equation}
In the modified constitutive equation we have a new constant
$\tau_r$ which is often called the ``relaxation time".
Correspondingly, the heat equation (\ref{heat2}) is also modified,
\begin{equation}
\tau_r\frac{\partial^2 T}{\partial t^2}+\frac{\partial T}{\partial t}=\kappa \frac{\partial^2 T}{\partial x_{i}^2}\,,\qquad c_s=\sqrt{\kappa/\tau_r}\,.
\label{heat3}
\end{equation}
In the literature the above equation is known as a telegraph
equation. While the original heat equation can be classified as a
parabolic equation, the telegraph equation belongs to the family
of hyperbolic equations. Causality
is not violated in Eq.~(\ref{heat3}) simply because we can now, by
choosing the relaxation time $\tau_r$ to be large, reduce the
signal velocity $c_s$ to values smaller than the speed
of light $c$.

In relativistic hydrodynamics the relaxation terms introduced
above can be obtained by modifying the entropy
current in the following way,
\begin{equation}
S^\mu=su^\mu+\mathcal{O}(\delta T^{\mu\nu})+\mathcal{O}\left((\delta T^{\mu\nu})^2\right)\,.
\end{equation}
By including the quadratic dissipative terms we are starting to
work within the framework of $2^\mathrm{nd}$ order theory. The
non-equilibrium entropy current vector $S^\mu$ in the
$2^\mathrm{nd}$ order theory has linear + quadratic dissipative
term(s) constructed from $(V^\mu,\Pi,\pi^{\mu\nu})$.
Again, we demand the 2$^\mathrm{nd}$ law of thermodynamics, $\partial_\mu S^{\mu}>0$.
Thus, quadratic dissipative terms modify the constitutive equations
which now read,
\begin{eqnarray}
\tau_\pi\Delta^{\mu\alpha}\Delta^{\nu\beta}\dot{\pi}_{\alpha\beta} + \pi^{\mu\nu} &=& 2\eta\left<\nabla^\mu u^\nu\right>+\cdots\,,\label{ce1}\\
\tau_\Pi\dot{\Pi}+\Pi&=&-\zeta\partial_\mu u^\mu+\cdots\,.\label{ce2}
\end{eqnarray}
When compared to the constitutive equations of the $1^\mathrm{st}$
order theory, (\ref{sst}) and (\ref{bp}), we see that in the
$2^\mathrm{nd}$ order theory in each constitutive equation
a relaxation term appears. Relaxation terms include $\tau_\pi$ and
$\tau_\Pi$, which are the relaxation times. It is important to
note that in the $2^\mathrm{nd}$ order theory the constitutive
equations are no longer algebraic equations. As a consequence,
dissipative currents become dynamical quantities like
thermodynamical variables. The constitutive equations with relaxation terms
have been employed in recent viscous fluid simulations
\cite{Muronga:2001zk,Muronga:2003ta,Muronga:2006zw,Muronga:2006zx,Heinz:2005bw,Song:2007fn,Baier:2006um,Baier:2006sr,Baier:2006gy,Romatschke:2007jx,Romatschke:2007mq,Dusling:2007gi,Koide:2006ef} 
\footnote{Some of the references here do not employ the same equations
as mentioned here.
There are still some hot debates
how to formulate the correct
\textit{relativistic} equation of viscous fluids
or which terms in the constitutive equations
of the 2$^{\mathrm{nd}}$ order theory should be kept
in the simulations.}.

Finally, we outline the Bjorken's equation in the $2^\mathrm{nd}$ order theory,
\begin{eqnarray}
\frac{de}{d\tau}&=&-\frac{e+P_s}{\tau}\left(1-\frac{\pi}{sT}+\frac{\Pi}{sT}\right)\,,\\
\tau_\pi\frac{d\pi}{d\tau}+\pi&=&\frac{4\eta}{3\tau}-\frac{\pi\tau_\pi}{2\tau}-\frac{\pi\eta T}{2}\frac{d}{d\tau}\frac{\tau_\pi}{\eta T}\,,\\
\tau_\Pi\frac{d\Pi}{d\tau}+\Pi&=&-\frac{\zeta}{\tau}-\frac{\Pi\tau_\Pi}{\tau}-\frac{\Pi\zeta T}{2}\frac{d}{d\tau}\frac{\tau_\Pi}{\zeta T}\,,
\end{eqnarray}
where
\begin{equation}
\pi=\pi^{00}-\pi^{zz}\,.\label{pi}
\end{equation}
It is easy to show that the above formulas reduce to the ones in the 1$^\mathrm{st}$
order theory if one takes $\tau_\pi \rightarrow 0$ and $\tau_\Pi \rightarrow 0$.
We remark here that, contrary to the $1^\mathrm{st}$
order theory, one needs to specify initial conditions for dissipative currents
in the $2^\mathrm{nd}$ order theory.

\subsection{Summary}

Let us summarize the main points so far:
\begin{itemize}
\item Hydrodynamics is a framework to describe the space-time evolution of matter under
local thermal equilibrium;
\item A na\"{\i}ve extension of Navier-Stokes equation to 
its relativistic version, which is called the first order theory,
has problems on instabilities and causality;
\item Relaxation terms are needed in the constitutive equations to resolve the above issues;
\item These terms naturally arise in the constitutive equations when
the $2^\mathrm{nd}$ order corrections of dissipative currents are considered in the entropy current.
\end{itemize}

\section{Applications}
\label{s:applications}
In this section we apply the formalism of hydrodynamics to heavy-ion collisions.
As already noted in Sec.~\ref{s:mainIntrodution},
we do not argue recent analyses in terms of viscous hydrodynamics. 
We show only results from ideal hydrodynamic models.
One can also consult recent other reviews of hydrodynamic models
at RHIC which complement the present lecture 
note~\cite{Kolb:2003dz,Huovinen:2003fa,Huovinen:2006jp,Hirano:2004ta,Hirano:2007gc,Hama:2004rr,Grassi:2004dz,Nonaka:2007nn}.
We start by discussing
some basic tests of whether the system produced in
H.I.C. can be described by thermodynamic quantities.
Then we discuss collective flow and introduce ideal hydrodynamic models to describe
the flow phenomena in H.I.C.
Finally we show results from ideal hydrodynamic models and compare them
with experimental data.

\subsection{Basic checks of observables at RHIC}
\label{s:basiccheck}

Recent lattice QCD results show \cite{Cheng:2007jq} the energy density as a function of the
temperature suddenly increases by $\sim 1$ GeV/fm$^3$
at the (pseudo-)critical temperature $T_c \sim 190$ 
MeV~\footnote{Energy density increases
with temperature rapidly but smoothly.
So this is not a phase transition but a cross over in a 
thermodynamically strict sense.
This is the reason why we call it pseudo-critical temperature here.}.
Above this temperature, the system is supposed to be
in the deconfined QGP.
The first check is whether the energy density produced in H.I.C.
is sufficient to form a QGP.
Phenomenologically, the energy density in H.I.C. can be estimated
through Bjorken's formula \cite{Bjorken:1982qr} \footnote{This formula neglects
the effect of $pdV$ work. If the system is kinetically equilibrated,
the energy density should be larger than the value obtained by
this formula \cite{Gyulassy:1983ub,Ruuskanen:1984wv}.
}
\begin{equation}
\epsilon_{\mathrm{Bj}}(\tau)=\frac{\left<m_T\right>}{\tau\pi R^2}\frac{dN}{dy}\,.
\label{eq:BJene}
\end{equation}
Here $\left<m_T\right>$ is the mean transverse mass, $y=\frac{1}{2}\ln\frac{E+p_z}{E-p_z}$ is the rapidity, 
$\frac{dN}{dy}$ is the number of
particles per unit rapidity,
$\tau = \sqrt{t^2-z^2}$ is the proper time and $R$ is an effective transverse radius.
The energy density obtained above depends on the proper time
since the system is supposed to expand in the longitudinal direction
with the expansion scalar $\theta=1/\tau$. 
One can compare Bjorken's energy density to the energy density from lattice QCD
simulations to see whether it is sufficient energy density to form a QGP.
\begin{figure}
\centering
\includegraphics[width=6cm]{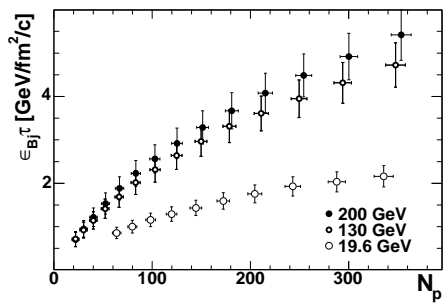}
\caption{$\varepsilon_{\mathrm{Bj}}\tau$ versus the number of participants
at three collision energies \cite{Adler:2004zn}.}
\label{fig:PHENIXdEdeta}
\end{figure}
Figure \ref{fig:PHENIXdEdeta} shows the PHENIX data
on $\varepsilon_{\mathrm{Bj}}\tau$ versus the number of participants
at three collision energies \cite{Adler:2004zn}.
If $\tau$ is taken to be 1 fm/$c$,
the Bjorken's energy densities
at $\sqrt{s_{NN}} = $ 130 and 200 GeV are well above the energy density
at the transition region $\sim 1$ GeV/fm$^3$.
Therefore sufficient energy is deposited
in the central rapidity region in H.I.C. at RHIC.
However, attention should be paid to the interpretation.
The above formula just counts the total measured energy divided by the volume
of a cylinder.
So the system is not necessary thermalized.
In this sense, this is a necessary condition, not a sufficient condition, to
form a QGP.

\begin{figure}
\centering
\includegraphics[width=1.0\linewidth]{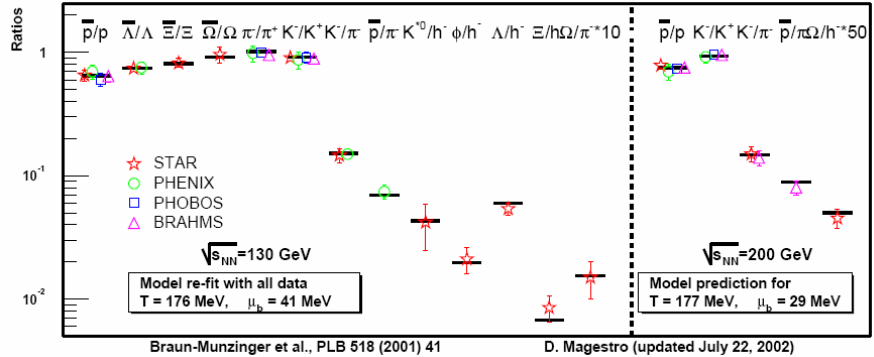}
\caption{Ratios of particle number at RHIC \cite{BraunMunzinger:2003zd}. }
\label{fig:slide47}
\end{figure}

The next basic check is whether the matter in H.I.C.
reaches chemical equilibrium.
Assuming thermal and chemical equilibrium,
we can calculate the number density of a certain
particle species
\begin{eqnarray}
n_i(T,\mu)&=&\frac{g}{2\pi^2}\int_0^\infty\frac{p^2dp}{\exp\left[(E_i-\mu_i)/T\right]\pm
1}\,.
\end{eqnarray}
$n_i$ gives the number density of particle species $i$ as a function of
the temperature $T$ and chemical potential $\mu_i$.
$g$ is the degeneracy of the particle,
$p$ is the momentum and $E$ is the energy.
We further assume the measured particle number is fixed
at a certain temperature and chemical potential, which is called chemical
freezeout.
Then the average number of particles, $\left<N_i\right>$,
can be estimated by summing contribution from
particles directly emitted from the
 system with volume $V$ and contribution from resonance decays
\begin{eqnarray}
\left<N_i\right>&=&V\left[n_i^{\mathrm{th}}(T,\mu)+\sum_R\Gamma_{R\rightarrow
i}n_R(T,\mu)\right]\,.
\end{eqnarray}
Here $n_i^{\mathrm{th}}$ and $n_R$ are 
the number density of directly emitted particle $i$ and resonance $R$, respectively.
$\Gamma_{R\rightarrow i}$ is the branching ratio of the resonance $R$
decaying into species $i$.
When one looks at ratios of two particle numbers, the volume $V$ is canceled out.
Thus the particle ratios depend only on two parameters: the temperature $T$
and the baryonic chemical potential $\mu_B$.
In Fig.~\ref{fig:slide47},
various combinations of the particle ratio 
observed at RHIC are fitted by two parameters \cite{BraunMunzinger:2003zd}.
We find a remarkably good fit to data with only these two parameter\footnote{There are some
additional parameters in the recent statistical models such as
excluded volume correction, strangeness suppression factor and so on
for a better description of the data.}.
At $\sqrt{s_{NN}} =$ 130 GeV, the temperature is fitted to be 176 MeV
which is close to the critical
temperature from lattice QCD calculations.
At the temperature, which we call chemical freezeout temperature $T^{\mathrm{ch}}$,
the system ceases to be in chemical equilibrium.
So we expect the system reaches chemical equilibrium above $T^{\mathrm{ch}}$.
Again, one has to keep in mind
that this is a necessary condition since
even in e$^+$e$^{-}$ or pp collisions
observed particle ratios are fitted reasonably well by using statistical models \cite{Becattini:1997rv}. 
See also discussions in, \textit{e.g.}, Refs.~\cite{Rischke:2001bt,Koch:2002uq}.

\begin{figure}
  \centerline{
    \begin{tabular}{cp{0.25cm}c}
      \subfigure[Isotropic case]{
        \includegraphics[height=6cm]{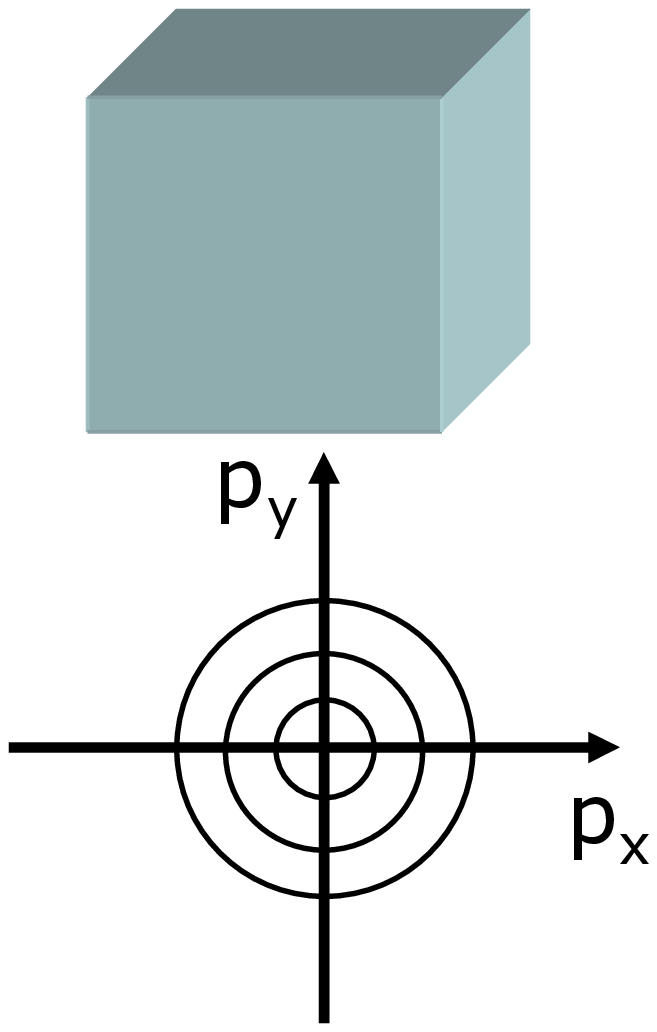}
\label{fig:slide49a}
      } & &
      \subfigure[Lorentz boosted in positive $x$ direction]{
        \includegraphics[height=6cm]{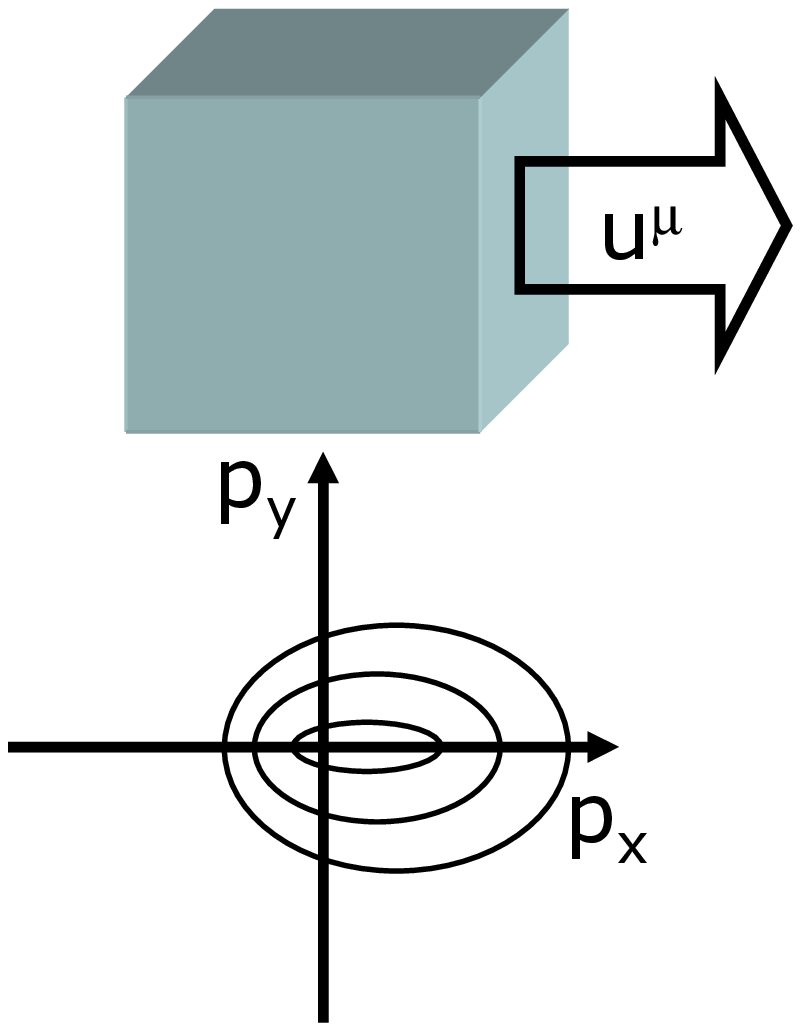}
\label{fig:slide49b}
      }
    \end{tabular}
  }
  \caption{Fluid elements at rest and at a finite velocity
  in $x$ direction. Momentum distribution in the latter case
  is distorted by Lorentz boost
  along $x$ axis.}
  \label{fig:slide49}
\end{figure}

\begin{figure}
\centering
\includegraphics[width=6cm]{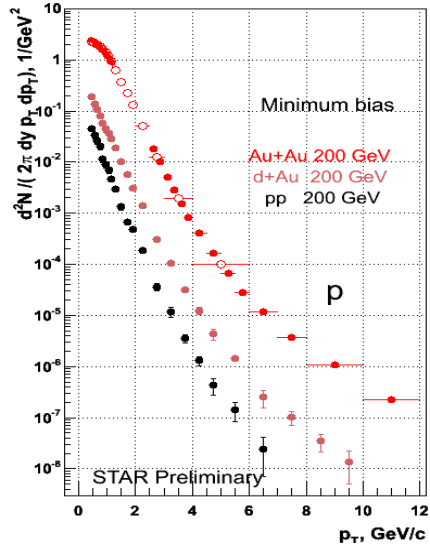}
\caption{Proton spectrum for pp (black), dAu (pink) and AuAu (red) collisions.
Adopted from a presentation file by O.~Barannikova at Quark Matter 2005, Budapest, Hungary \cite{Barannikova:2006dk}.}
\label{fig:slide51}
\end{figure}

\begin{figure}
\centering
\includegraphics[width=6cm]{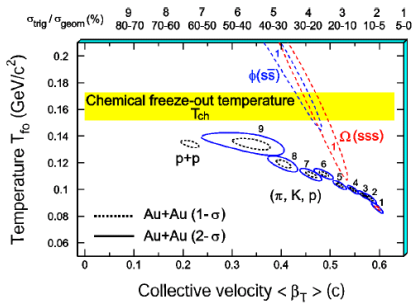}
\caption{Fitted parameters in blast wave model calculations \cite{Adams:2005dq}.}
\label{fig:slide52}
\end{figure}

The last basic check is whether the matter reaches kinetic equilibrium.
If we suppose a system in H.I.C. is in kinetic equilibrium,
the pressure is built
inside the system.
The matter is surrounded by vacuum, 
so pressure gradient in outward directions
generates collective flow and, in turn,
the system expands radially.
The momentum distribution in kinetically equilibrated matter
is isotropic.
On the other hand, when the matter is moving at a finite velocity
the momentum distribution is
Lorentz boosted.
This is illustrated in Fig.~\ref{fig:slide49}.
If this kind of distortion in momentum distribution
can be observed experimentally,
one can obtain some information about kinetic equilibrium.
Assuming each fluid element expands radially at radial flow velocity $v_{T}$,
the $p_T$ spectra for pions and protons
can be calculated by convoluting these distorted
momentum distributions over azimuthal direction (blast wave model \cite{Siemens:1978pb,Schnedermann:1993ws}).
 Here $p_{T}$ 
is the transverse momentum which is perpendicular to
the collision axis.
The green curves are results
with $T=100$ MeV and radial flow velocity $v_{T} = 0.5$.
On the other hand, the red curves are results
with $T=160$ MeV and vanishing flow $v_{T} = 0$.
For light particles like pions,
there is almost no sensitivity to distinguish the two cases:
Reduction of temperature is almost compensated by radial flow.
However, in the case
of heavier particles like protons, a clear difference can be
seen between these two cases:
There is a shoulder structure at low $p_T$
resulting from radial flow.
\begin{figure}
\centering
\includegraphics[width=6cm]{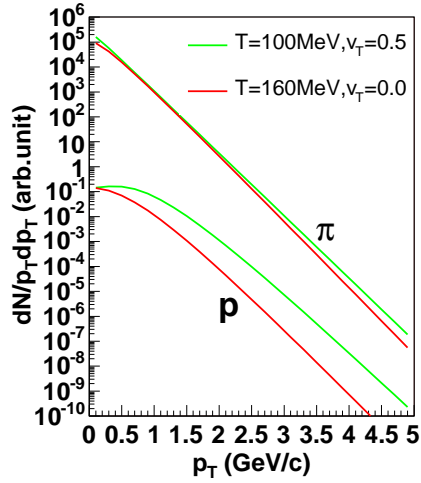}
\caption{$p_{T}$ spectra for pions and protons from a thermal plus boost picture.
See text for details.}
\label{fig:slide50b}
\end{figure}
This kind of spectral change is observed in H.I.C.,
as can be seen in Fig.~\ref{fig:slide51}.
It shows the proton $p_{T}$ spectra for p+p (black), d+Au (pink) and Au+Au (red) collisions
obtained by STAR Collaboration \cite{Barannikova:2006dk}.
For p+p and d+Au collisions the spectra have just a power-law shape.
However, in Au+Au collisions, one sees a shoulder structure at low $p_T$ ($<1$ GeV/$c$).
This is consistent with a thermal plus boost picture and suggests that
a large pressure could be built up in Au+Au collisions.
One can fit the $p_{T}$ spectrum using a blast wave parametrization \cite{Siemens:1978pb,Schnedermann:1993ws} and
obtains decoupling temperature $T^{\mathrm{dec}}$
and the mean collective flow velocity as a function of the centrality.
Even for pp collisions these parameters are finite
(see Fig.~\ref{fig:slide52}) \cite{Adams:2005dq}, which indicates that a more
sophisticated model would be needed to interpret the
data.
This kind of spectral change can also be seen in results from
kinetic theories in which kinetic equilibrium
is not fully achieved.
Therefore it is indispensable to perform a systematic study based on a
more sophisticated dynamical framework.

We have obtained the necessary conditions for studying the QGP:
(1) The energy density can be well above the critical value which is
predicted from lattice QCD simulations; 
(2) A chemical freezeout temperature extracted from particle ratios
is close to pseudo-critical temperature which is again from lattice
QCD simulations;
(3) High pressure can be built up in H.I.C., which suggests 
the system reaches kinetic equilibrium.
If one of them was not confirmed through these basic checks,
one would not need to go next steps towards
detailed studies of the QGP in H.I.C.

\subsection{Elliptic flow}
\label{s:ellipticflow}

\begin{figure}
  \centerline{
    \begin{tabular}{cp{0.25cm}c}
      \subfigure[In the reaction plane]{
        \includegraphics[width=\subfigsize]{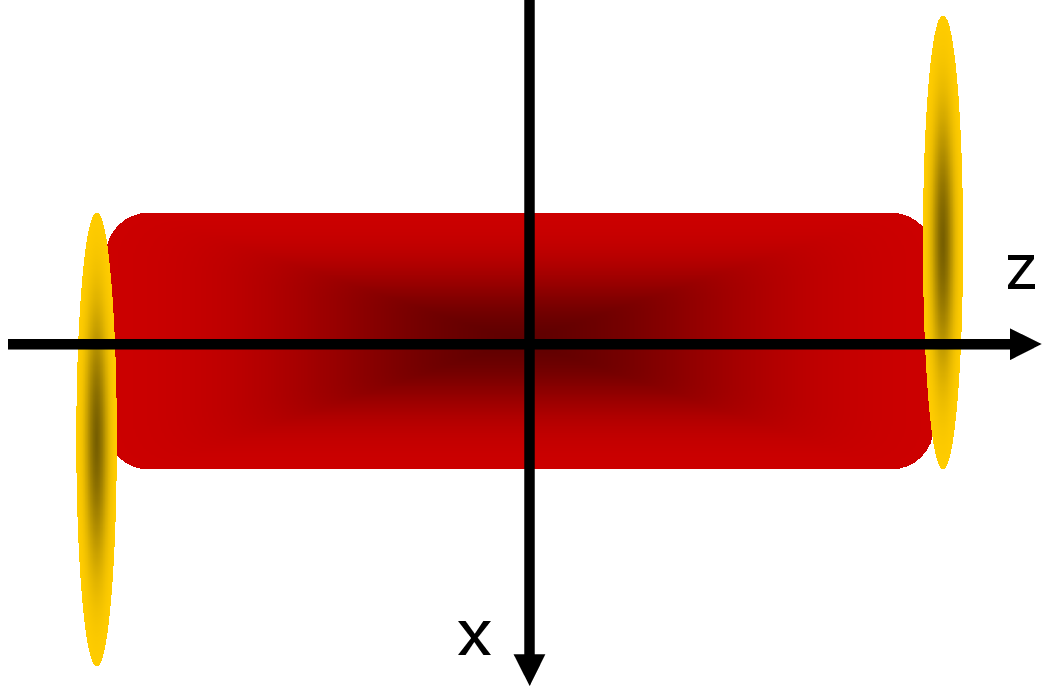}
        \label{fig:slide55a}
      } & &
      \subfigure[In the transverse plane]{
        \includegraphics[width=\subfigsize]{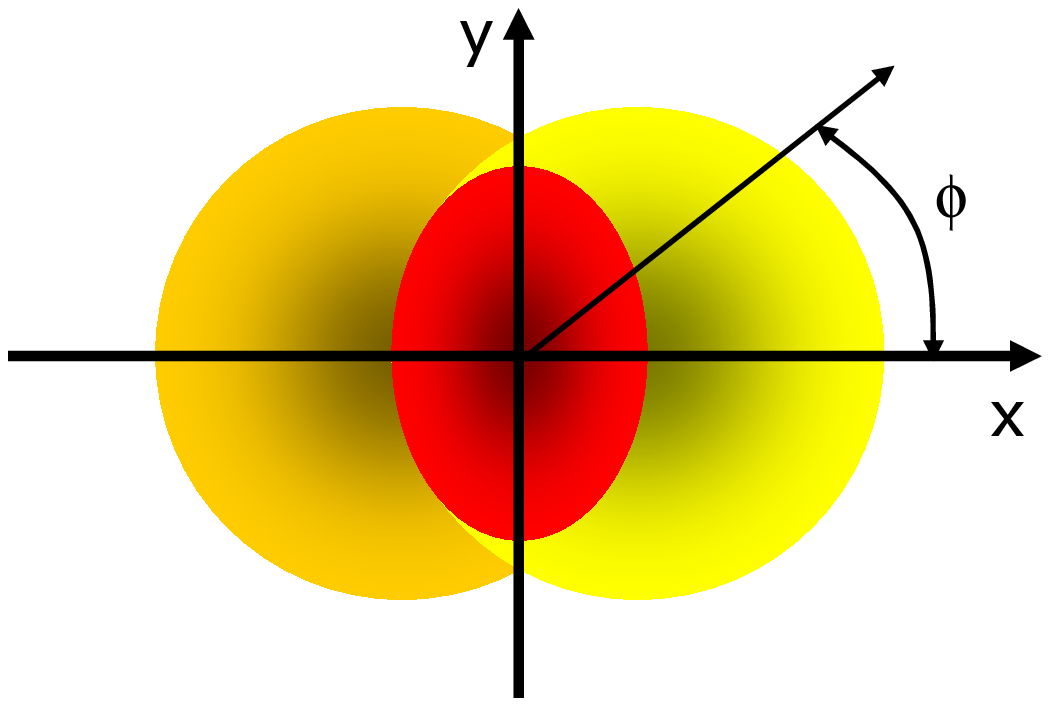}
        \label{fig:slide55b}
      }
    \end{tabular}
  }
  \caption{Illustration of a H.I.C.}
  \label{fig:slide55}
\end{figure}

\begin{figure}
  \centerline{
    \begin{tabular}{cp{0.25cm}c}
      \subfigure[First harmonic $v_1$]{
        \includegraphics[width=\subfigsize]{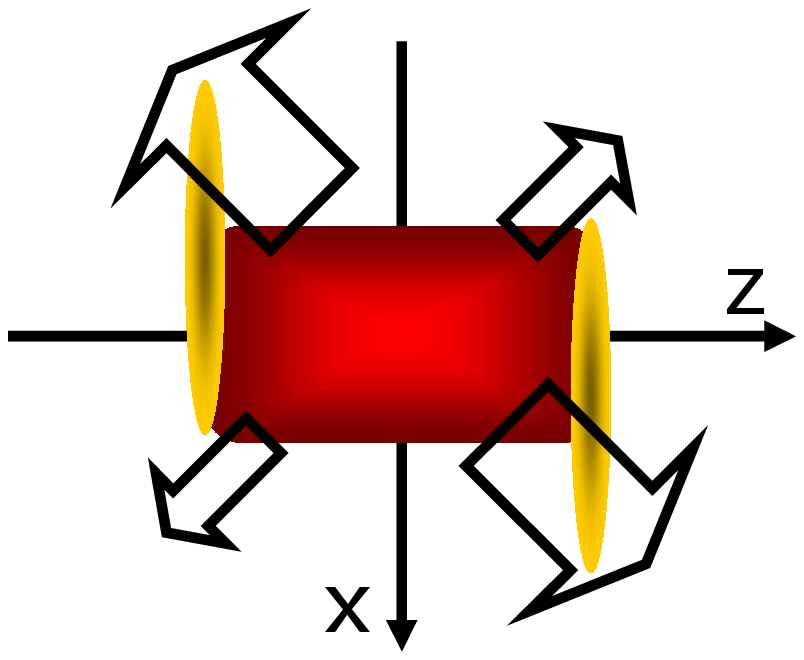}
        \label{fig:slide56a}
      } & &
      \subfigure[Second harmonic $v_2$]{
        \includegraphics[width=\subfigsize]{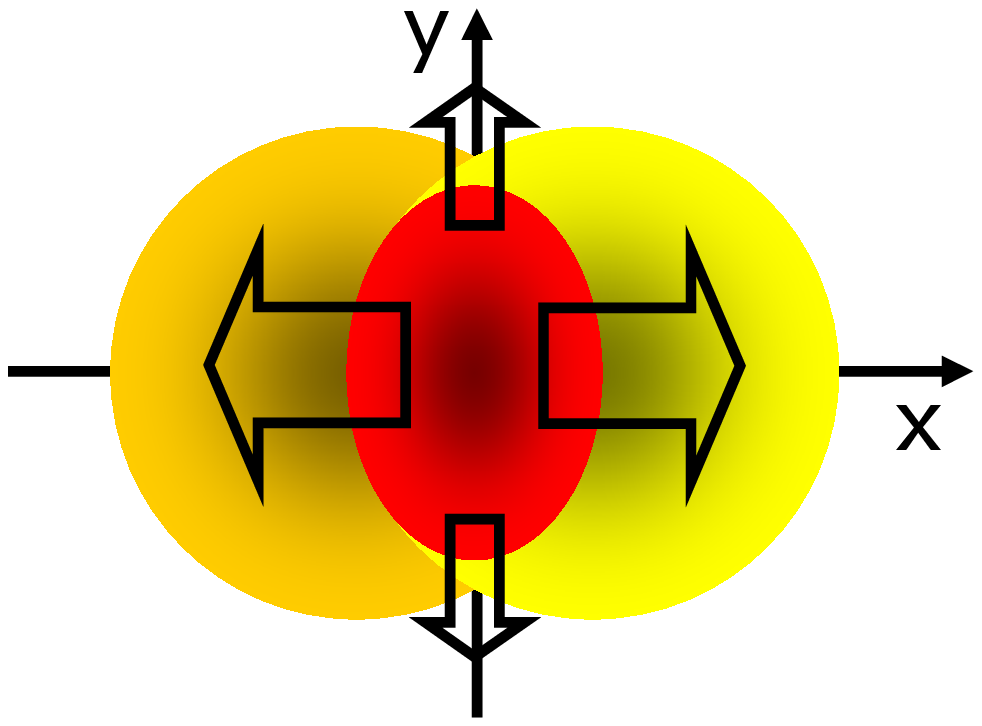}
        \label{fig:slide56b}
      }
    \end{tabular}
  }
  \caption{Anisotropic transverse flow}
  \label{fig:slide56}
\end{figure}

Before going to a detailed discussion on 
the hydrodynamic models, we discuss collective flow, in particular, anisotropic
transverse flow.
Here ``collective flow" is meant by the correlation between
position of matter and direction of flow, which is not necessary to
be hydrodynamically evolving matter.
A good example has already appeared in the previous subsection.
In the case of radial flow, velocity of expanding matter
has a component parallel to the radial coordinate.
Figure \ref{fig:slide55} shows a heavy ion collision in the reaction plane (left)
 and
transverse plane (right).
In such a collision a region of the locally equilibrated state can be created.
In the transverse plane the overlap region has an almond like shape,
so the region is anisotropic with respect to the azimuthal angle.
The azimuthal momentum distribution can be expanded into a Fourier series
\footnote{Here we suppose azimuthal angle is measured from
reaction plane. Of course,
in the experimental situations, the reaction plane is not known a priori.
We will not go into details of how to find reaction plane experimentally.}
\begin{eqnarray}
\frac{dN}{d\phi} & = & \frac{N}{2\pi}\left[1+2v_1\cos(\phi)+2v_2\cos(2\phi)+\cdots\right]\,,\\
v_n & = & \frac{\int d\phi\cos(n\phi)\frac{dN}{d\phi}}{\int
d\phi\,\frac{dN}{d\phi}}=\left<\cos(n\phi)\right>\,.
\end{eqnarray}
where $\phi$ is the azimuthal angle of momentum and $v_n$ are the Fourier coefficients
of $n$-th harmonics  \cite{Poskanzer:1998yz}.
Because of the symmetry around the $y$-axis the sine terms vanish.
The first and the second harmonics, $v_1$ and $v_2$,
are called directed and elliptic flow parameters, respectively.
The first harmonic, $v_1$, is illustrated in Fig.~\ref{fig:slide56a}.
Particles are emitted preferably, \textit{e.g.}, in the direction of the large arrows
in the reaction plane.
Directed flow is significantly seen
near the beam rapidity region, but 
vanishes near midrapidity due to symmetry of the collision geometry.
The second harmonic, $v_2$, is much more relevant for studying matter around midrapidity
in H.I.C. at relativistic energies 
since spectators already fly away \cite{Ollitrault:1992bk},
therefore a lot of efforts to measure
$v_2$ have been made at RHIC so far.
One of the first observables was actually $v_2$ measured by STAR
Collaboration \cite{Ackermann:2000tr}.
It is illustrated in Fig.~\ref{fig:slide56b}.

\begin{figure}
  \centerline{
    \begin{tabular}{cp{0.25cm}c}
      \subfigure[Large mean free path]{
        \includegraphics[height=6cm]{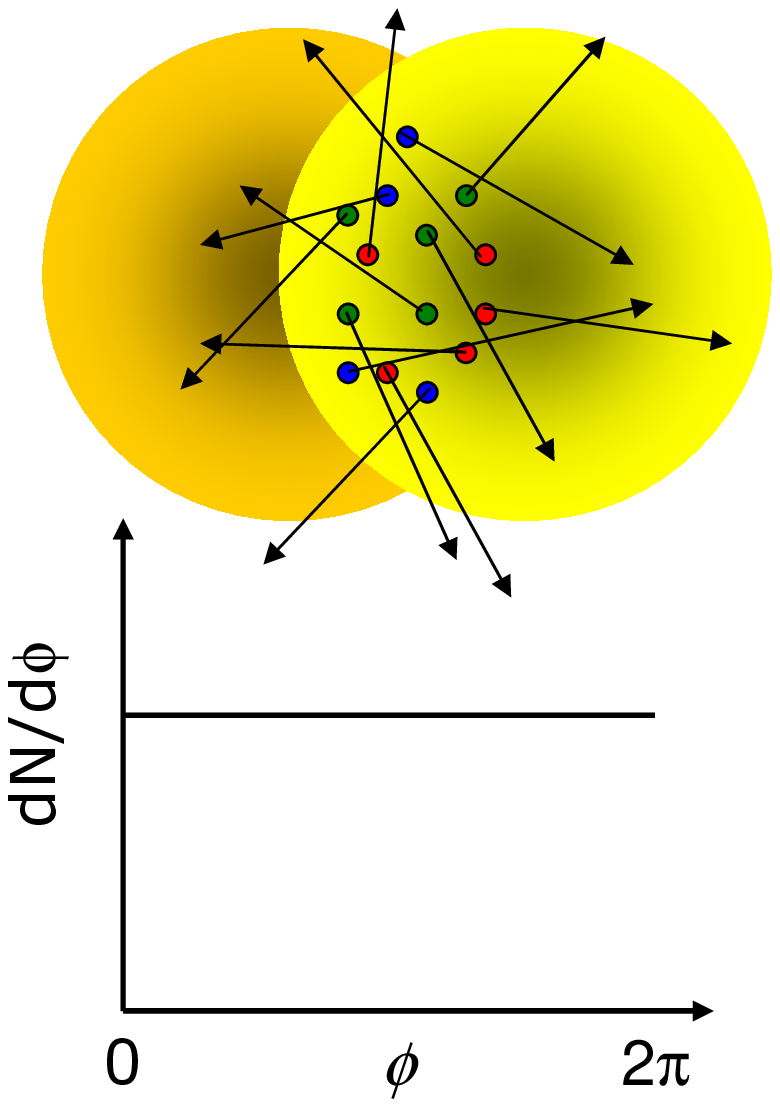}
        \label{fig:slide57ad}
      } & &
      \subfigure[Small mean free path]{
        \includegraphics[height=6cm]{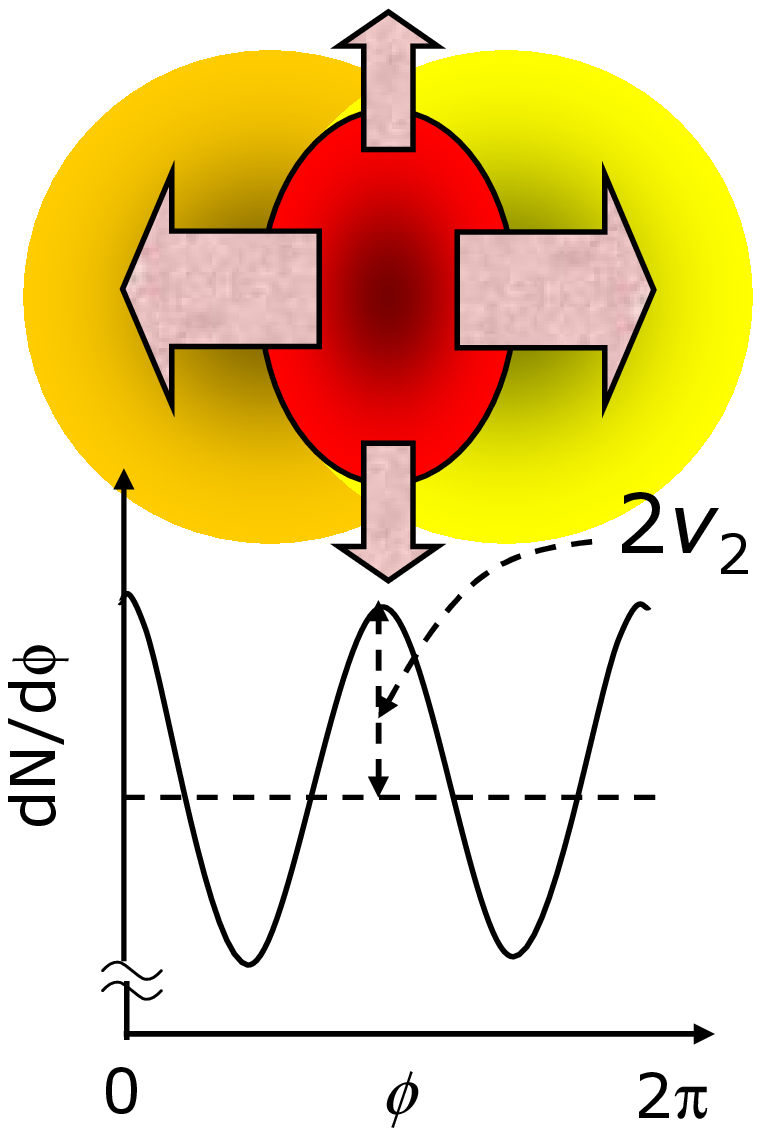}
        \label{fig:slide57ce}
      }
    \end{tabular}
  }
  \caption{Normalized azimuthal distribution $dN/d\phi$ of a non-central H.I.C.}
  \label{fig:slide57}
\end{figure}

Elliptic flow is how the system responds to
the initial spatial anisotropy \cite{Ollitrault:1992bk,Heiselberg:1998es,Sorge:1998mk,Voloshin:1999gs}.
Suppose two extreme situations illustrated in Fig.~\ref{fig:slide57}.
In the first case (see Fig.~\ref{fig:slide57ad})
the mean free path among the produced particles is much larger than the typical
size of the system.
In this case the azimuthal distribution of particles does not depend on
azimuthal angle on average due to the symmetry of the
production process.
The other extreme case is when the mean free path is very small compared to the
typical system size (see Fig.~\ref{fig:slide57ce}).
In this case hydrodynamics can be applied to describe the 
space-time evolution of the system.
The pressure gradient along the horizontal axis 
is much larger than along the vertical axis due to the geometry.
So the collective flow is enhanced along the horizontal axis
rather than along the vertical axis
and, in turn, the azimuthal distribution gets oscillated.
The amplitude of this oscillation 
in the normalized azimuthal distribution
describes exactly the elliptic flow parameter.
In this way, the elliptic flow is generated by the spatial anisotropy of the almond shape
due to multiple interactions among the produced particles.
We have good opportunities to extract some information about the mean free path from
the elliptic flow analysis.

The eccentricity is a very important quantity to interpret elliptic flow phenomena.
To quantify the initial almond shape, the following formula can be used
\begin{equation}
\varepsilon=\frac{\left<y^2-x^2\right>}{\left<y^2+x^2\right>}\,.
\label{eq:stdecc}
\end{equation}
The brackets denote an average over the transverse plane 
with the number density of participants as a weighting function
\begin{equation}
\left<\cdots\right>=\int dxdy\cdots n_{\mathrm{part}}(x,y)\,.
\end{equation}
This is sometimes called the standard eccentricity.
If the system is elongated along the $y$-axis,
the eccentricity is positive.
In more realistic situations, 
the eccentricity fluctuates from event to event.
This fluctuation of the initial eccentricity 
\cite{Miller:2003kd,Zhu:2005qa,Drescher:2006ca,Andrade:2006yh,Broniowski:2007ft,Alver:2008zz}
 is important 
to understand the elliptic flow
in the small system such as Cu+Cu collisions or peripheral Au+Au collisions.
\begin{figure}
\centering
\includegraphics[height=4cm]{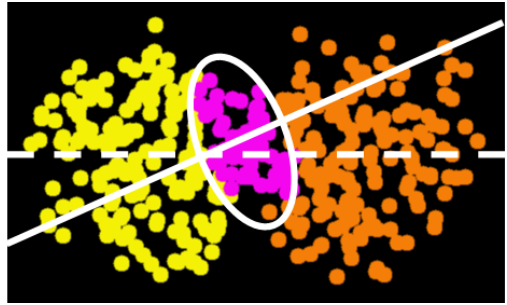}
\caption{An example of participants (magenta) and spectators (yellow and orange)
in a H.I.C. from a Monte Carlo Glauber model. Adopted from a presentation file
by D.~Hofman at Quark Matter 2006, Shanghai, China.}
\label{fig:slide59}
\end{figure}
Figure~\ref{fig:slide59} shows a sample event
projected into the transverse plane
from a Monte Carlo Glauber model.
Participants are shown in magenta and spectators are in yellow and orange.
In this case one could misidentify the tilted line as the reaction plane,
while the true reaction plane is the horizontal axis (dashed line).
The angle of the tilted plane with respect to the true reaction plane
fluctuates event by event.
Of course we cannot observe the true reaction plane from experimental data.
On the other hand, an apparent reaction plane (tilted line in Fig.~\ref{fig:slide59}) is determined
also by elliptic flow signal itself.
Another definition, called the participant eccentricity, is much more relevant
for quantifying the almond shape in the event by event basis
\begin{eqnarray}
\varepsilon_{\mathrm{part}} & = & \frac{\sqrt{(\sigma_y^2-\sigma_x^2)^2+4\sigma_{xy}^{2}}}
{\sigma_x^2 + \sigma_y^2}\,,\\
\sigma_{x}^2 & =  &\left\{ x^2 \right\} - \left\{x\right\}^2\,,\\
\sigma_{y}^2 & =  &\left\{ y^2 \right\} - \left\{y\right\}^2\,,\\
\sigma_{xy} & =  &\left\{ xy \right\} - \left\{x\right\}\left\{y\right\}\,.
\end{eqnarray}
Now the average $\left\{\cdots \right\}$ is taken over in a single event
generated by a Monte Carlo Glauber Model.

\begin{figure}
\centering
\includegraphics[width=6cm]{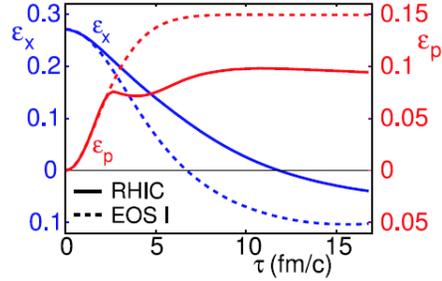}
\caption{The spatial eccentricity $\varepsilon_x$ and the momentum eccentricity $\varepsilon_p$
as a function of the proper time $\tau$ 
in Au+Au collisions at $b=7$ fm \cite{Kolb:2003dz}.
Solid and dashed curves correspond to two different sets
of the EoS.}
\label{fig:slide60a}
\end{figure}

In the following, the important 
properties of elliptic flow are demonstrated through
hydrodynamic/transport simulations of H.I.C.
In hydrodynamic simulations, the eccentricity is usually defined by
weighting local energy density $e(x,y)$
or local entropy density $s(x,y)$ in the transverse plane
rather than the number density of participants $n_{\mathrm{part}}(x,y)$.
Figure~\ref{fig:slide60a} shows the eccentricity
$\varepsilon_x$ and the momentum eccentricity 
\begin{equation}
\varepsilon_p = \frac{\int dxdy (T_0^{xx}-T_0^{yy})}{\int dxdy (T_0^{xx}+T_0^{yy})}
\end{equation}
as a function of the proper time from a hydrodynamic simulation
assuming Bjorken scaling solution in the longitudinal direction 
and two different sets of the EoS \cite{Kolb:2003dz}.
Details of hydrodynamic models will be discussed later.
\begin{figure}
\centering
\includegraphics[width=6cm]{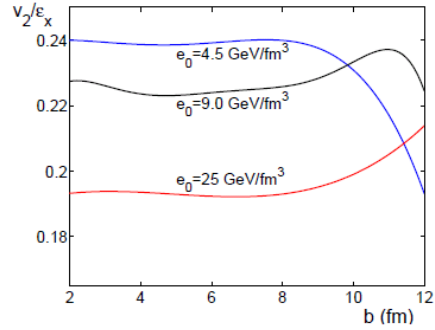}
\caption{$v_2/\varepsilon_x$ as a function of impact parameter $b$ \cite{Kolb:2000sd}.}
\label{fig:slide60b}
\end{figure}
The spatial eccentricity $\varepsilon_x$
decreases as the system expands and the momentum anisotropy
rapidly increases at the same time.
So the spatial anisotropy turns into the momentum anisotropy.
The momentum anisotropy $\varepsilon_p$ is created and saturates in the first several femtometers,
so the observed $v_2$ is expected to be sensitive to the initial stage of the
collision. 
Figure~\ref{fig:slide60b} shows 
the impact parameter dependence of the ratio of output ($v_2$) to input ($\varepsilon_x$) \cite{Kolb:2000sd}
which can be understood as a response of the system.
Ideal hydrodynamics predicts that $v_2$ is roughly proportional to the eccentricity
\begin{equation}
v_2\approx 0.2\varepsilon\,.
\end{equation}

\begin{figure}
\centering
\includegraphics[width=6cm]{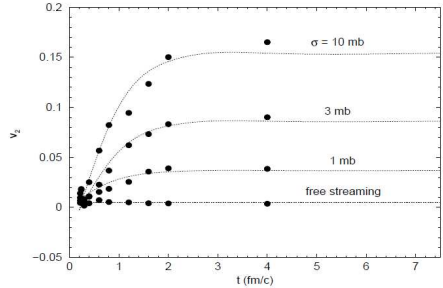}
\caption{$v_2$ as a function of proper time from Boltzmann calculations for
different gluon cross sections \cite{Zhang:1999rs}. Curves are guide to eyes.}
\label{fig:slide61}
\end{figure}

Figure~\ref{fig:slide61} shows a result from a kinetic approach based on
the Boltzmann equation for gluons undergoing elastic scattering only \cite{Zhang:1999rs}
\footnote{Inelastic scattering ($gg \leftrightarrow ggg$)
is implemented in a kinetic approach only recently.
Although this is a higher order process in perturbative expansion, 
it turns out to affect elliptic flow significantly. See Refs.~\cite{Xu:2004mz,Xu:2007aa,Xu:2007ns}}.
Starting with a uniform distribution in an almond shape in coordinate space
and thermal distribution in momentum space,
the multi-gluon system expands according to the Boltzmann equation
with various transport cross sections\footnote{
In kinetic theories, momentum exchanges among particles
are responsible for equilibration.
However, forward scattering with very small scattering angle
is insufficient for the system to equilibrate. So the effective (transport) cross section
can be defined as $\sigma_{\mathrm{tr}}=\int  d\theta_{\mathrm{cm}} \sin^2 \theta_{\mathrm{cm}}
\frac{d\sigma}{d\theta_{\mathrm{cm}}}$,
where $\theta_{\mathrm{cm}}$ is scattering angle in the center of mass system between two 
scattering particles.}.
From this figure we can understand several important features
of the elliptic flow:
\begin{enumerate}
\item $v_2$ is not generated in the free streaming case,
so elliptic flow is generated indeed through secondary collisions;
\item Elliptic flow is generated
in the early stage of the collision and saturates after the first
2 to 3 fm/$c$;
\item The saturated value of $v_2$ is sensitive to the cross
section among the particles
\begin{equation}
\sigma_{\mathrm{tr}} \propto \frac{1}{\lambda} \propto \frac{1}{\eta}\,,
\end{equation}
where $\lambda$ is the mean free path and $\eta$ is the shear viscosity
calculated in the kinetic theory of gases;
\item In the limit of large
transport cross sections (strongly interacting limit),
the system is expected to reach 
the ideal hydrodynamic result\footnote{The Boltzmann equation is applied
to \textit{dilute} gases where two particle correlation can be ignored.
So one should keep in mind the applicability condition of the kinetic theory
in this case.} since $\eta \rightarrow 0$.
\end{enumerate}
Through measurement of $v_2$ and its analysis in terms of hydrodynamic/transport models,
one can extract the transport properties of the matter 
produced in H.I.C. In the next subsection, we discuss hydrodynamic modeling of H.I.C.

\subsection{Ideal hydrodynamic model}

Hydrodynamics introduced in Sec.~\ref{s:formalism}
is a general framework to describe the space-time evolution
of locally thermalized matter for a given equation of state (EoS).
This framework has been applied to the intermediate stage in H.I.C.
In this section, we neglect the effects of dissipation
and concentrate on discussion about ideal hydrodynamic models.
The main ingredient in ideal hydrodynamic models in H.I.C.
is the EoS
of hot and dense matter governed by QCD.
In addition, one also needs to assign initial conditions
to the hydrodynamic equations.
Hydrodynamics can be applied to a
system in which local thermalization is maintained.
However, in the final
state of H.I.C. the particles are freely streaming toward the detectors and
their mean free path is almost infinite.
This is obviously beyond the applicability of hydrodynamics. 
Hence we also need a 
description to decouple the particles from the rest of the system.
To summarize, the hydrodynamic modeling of H.I.C. needs an EoS, initial
conditions and a decoupling prescription. Modeling of
these ingredients in hydrodynamic simulations
has been sophisticated for these years and tested against a vast body of
RHIC data.

\begin{figure}
\centering
\includegraphics[width=6cm]{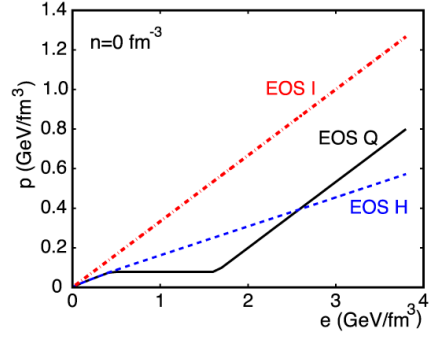}
\caption{Some typical EoS in hydrodynamic models~\cite{Kolb:2003dz}.}
\label{fig:slide64}
\end{figure}

We first look at
the EoS in more detail.
The EoS is in principle calculated from lattice QCD simulations.
The realistic results with (almost) physical quark masses 
are obtained recently \cite{Cheng:2007jq}.
However, if one wants to utilize the EoS from lattice simulations,
one needs to interpret the EoS in term of a hadron picture \cite{Karsch:2003vd}
since one calculates
momentum distributions of hadrons in the final decoupling stage.
For this purpose, the lattice EoS is compared with the resonance gas model
below $T_c$. If there exists a deviation between them,
it prevents ones from utilizing the lattice EoS directly
in hydrodynamic simulations.
Instead, in hydrodynamic simulations, the models of EoS depicted in
Fig.~\ref{fig:slide64} are conventionally used \cite{Kolb:2003dz}.
The most simple EoS (EOS I) 
is $P = e/3$ for an ideal gas of relativistic massless particles\footnote{
This EoS is always obtained in relativistic conformal field theories
in which the trace of energy-momentum tensor is 
vanishing $T^{\mu}_{\enskip \mu} = e - 3P = 0$.
So the particles are not necessarily ``free".}.
A more realistic EoS (EOS Q) includes the effect of hadron masses
and phase transition between hadronic matter and the QGP.
At low energy density the EoS is described by a hadron resonance gas model
(EOS H). This particular model includes almost all the hadrons in the Particle Data Table \cite{PDG},
while some models include only ground states of hadron multiplets or several low mass
resonances.
At high energy density, the EoS can be described by a bag model
\begin{equation}
P = \frac{1}{3}(e-4B)\,.
\end{equation}
The bag constant $B$ is tuned to 
match pressure of the QGP phase to that of a hadron resonance gas
at critical temperature $T_c$: $P_{\mathrm{QGP}}(T_c) = P_{\mathrm{hadron}}(T_c)$.
As discussed in Sec.~\ref{s:basiccheck},
a hadron gas in 
H.I.C. is not in chemical equilibrium below the chemical freezeout
temperature. $T^{\mathrm{ch}}$ which is closed to $T_c$, so the hadron phase
may not be chemically equilibrated in H.I.C.
A chemically frozen hadron resonance gas can be described
by introducing the chemical potential for each hadron \cite{Hirano02,Bebie,spherio,Qian,Teaney:2002aj,Kolb:2002ve,Huovinen:2007xh}.
The numbers $\tilde{N}_{i}$ including
all decay contributions from higher-lying resonances, 
$\tilde{N}_{i}=N_{i}+\sum_{R}b_{R\rightarrow iX}N_{R}$,
are conserved during the evolution in co-moving frame of fluid elements.
Here $N_{i}$ is the number of the $i$-th hadronic species in a fluid element
and $b_{R\rightarrow iX}$ is the effective branching ratio (a product
of branching ratio and degeneracy) of a decay process $R\rightarrow i+X$.
One can calculate the chemical potential as a function of temperature from the
following conditions:
\begin{eqnarray}
\frac{\tilde{n}_i(T, \mu_i)}{s(T, \{\mu_i \})} & = & \frac{\tilde{n}_i(T_c,  \mu_i=0 )}{s(T_c,  \{\mu_i \}=0)}\,.
\label{eq:PCE}
\end{eqnarray}
Instead of solving continuity equations for each hadron,
the effect of hadron number conservation can be embedded in the EoS of resonance gas
through $\mu_i(T)$ obtained above.

For a decoupling prescription, 
the Cooper-Frye formula \cite{Cooper:1974mv} is almost a unique choice to convert the hydrodynamic
picture to the particle picture
\begin{eqnarray}
E\frac{dN}{d^3p} & = & \int_\Sigma f(x, p, t) p\cdot
d\sigma(x)\,\\
& = &\frac{d}{(2\pi)^3}\int_\Sigma\frac{p\cdot
d\sigma(x)}{\exp\left[(p \cdot u(x)-\mu(x))/T(x)\right]\pm 1}\,,
\label{eq:CF}
\end{eqnarray}
where $E$ is the energy, $f$ is the phase space distribution,
$d$ the degeneracy of the 
particle under consideration
(\textit{e.g.}, $d=3$ for pions),
$p$ is the momentum, $d\sigma$ is
the normal vector to the freezeout surface element, $u$ is the four-velocity,
$\mu$ is the chemical potential and $T$ is the decoupling temperature
assuming isothermal freezeout hypersurface $\Sigma$.
Contribution from resonance decays should be taken into
account by applying some decay kinematics to the outcome of the
Cooper-Frye formula.
The decoupling temperature $T^{\mathrm{dec}}$
is fixed through \textit{simultaneous} fitting of
$p_{T}$ spectra for various hadrons in the low $p_{T}$ region.
In the blast wave model, decoupling temperature and radial flow
velocity are independent parameters to fit $p_{T}$ spectra.
On the other hand, there is a negative correlation between $T^{\mathrm{dec}}$
and average radial flow velocity in the hydrodynamic model: the lower decoupling temperature,
the larger average radial flow velocity.
This formula ensures the energy-momentum conservation on freezeout 
hypersurface $\Sigma$
as long as the EoS is calculated using the same distribution function.
If one puts resonances up to the mass of 2 GeV
in the resonance gas model, one should calculate all the contribution of
hadrons in the EoS. Otherwise, neglect of the contribution leads to violation of the energy momentum 
conservation\footnote{If the lattice EoS below $T_c$
cannot be described by a resonance gas model,
the Cooper-Frye formula violates the energy-momentum conservation on $\Sigma$.
This is the reason why
there are only few serious attempts of lattice EoS
to hydrodynamic simulations.}.
It should be noted that $p\cdot d\sigma$ term 
in Eq.~(\ref{eq:CF}) can be negative. This means the in-coming particles through $\Sigma$
are counted as a negative number. Although this seems peculiar,
this negative contribution is needed for global energy momentum conservation.

\begin{figure}
  \centerline{
    \begin{tabular}{cp{0.5cm}c}
      \subfigure[Sudden freezeout]{
        \includegraphics[width=\subfigsize]{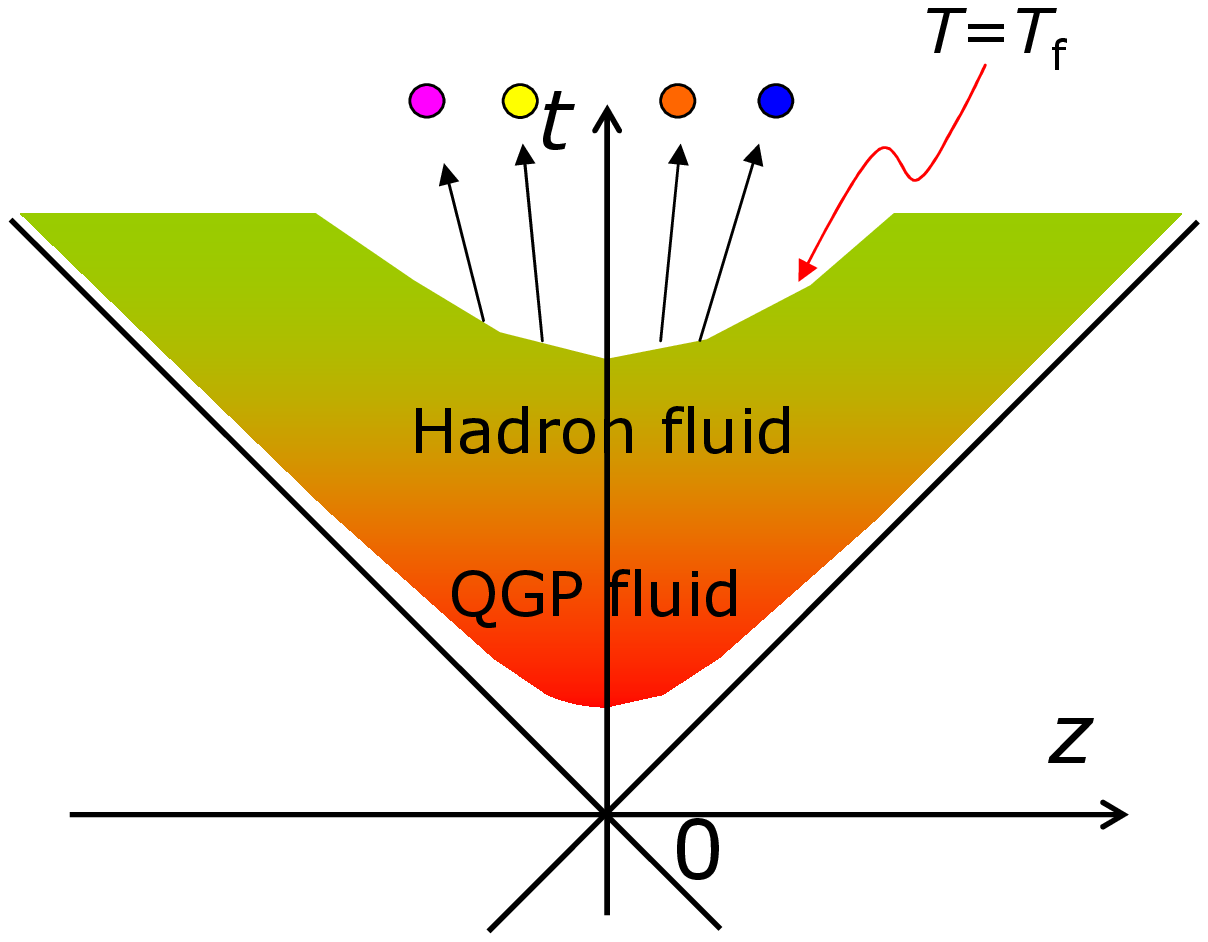}
        \label{fig:slide68a}
      } & &
      \subfigure[Gradual freezeout]{
        \includegraphics[width=\subfigsize]{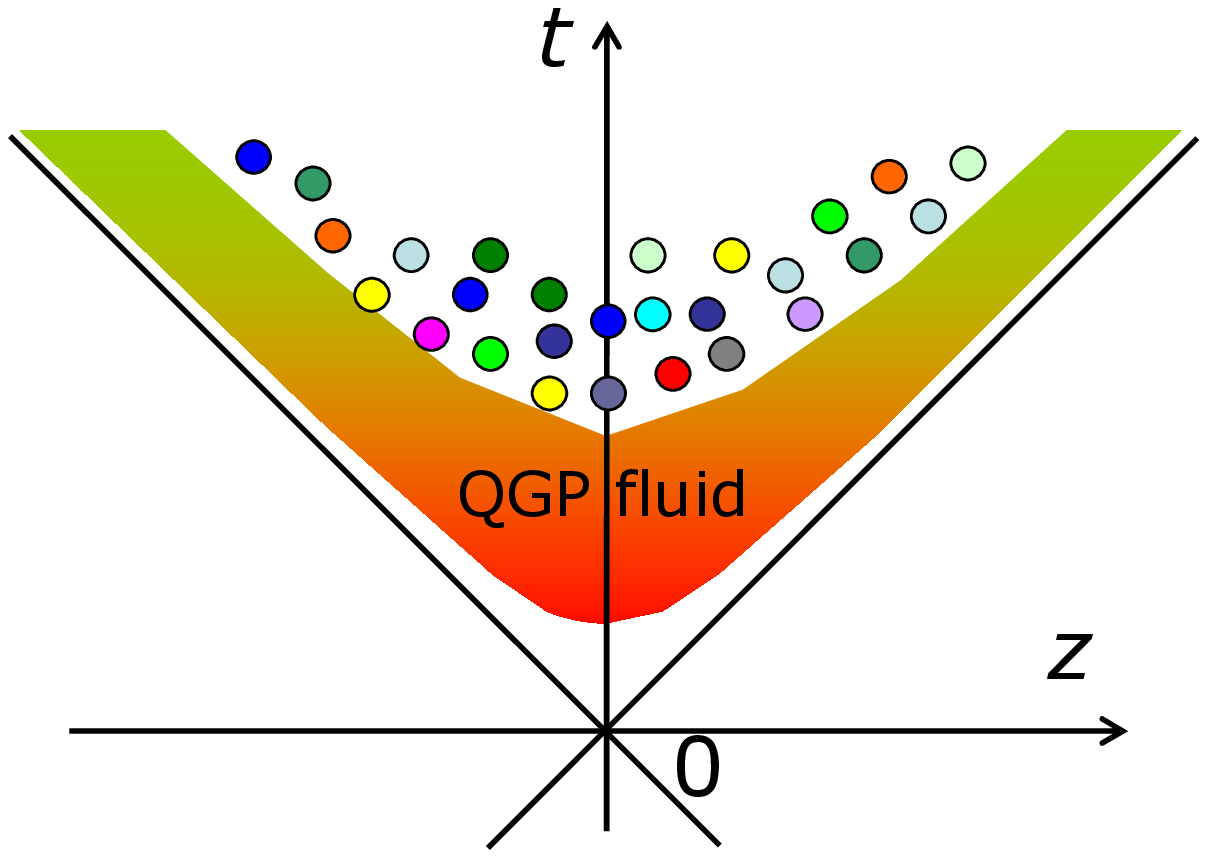}
        \label{fig:slide68b}
      }
    \end{tabular}
  }
  \caption{Two freezeout pictures in H.I.C.}
  \label{fig:slide68}
\end{figure}

The prescription to calculate the momentum distribution as above
is sometimes called the sudden freezeout model
since the mean free path of the particles changes
from zero (ideal fluid) to infinity (free streaming) within a thin layer $\Sigma$.
Although this model is too simple, it has been used in hydrodynamic
calculations for a long time.
It is illustrated in Fig.~\ref{fig:slide68a}.
Recently one utilizes hadronic cascade models to describe
the gradual freezeout \cite{Bass:2000ib,Teaney:2000cw,Hirano:2005xf,Hirano:2007ei,Nonaka:2006yn}.
As will be shown, this hadronic afterburner is mandatory in understanding
$v_2$ data.
Phase space distributions for hadrons
are initialized below $T_c$ by using the Cooper-Frye formula.
The hadronic cascade models describe
the space-time evolution of the hadron gas. This model is illustrated in
Fig.~\ref{fig:slide68b}.
This kind of hybrid approaches in which the QGP fluids are 
followed by hadronic cascade models
automatically describes both the chemical and
thermal freezeout and is much more realistic especially for the late stage.

\begin{figure}
\centering
\includegraphics[height=4cm]{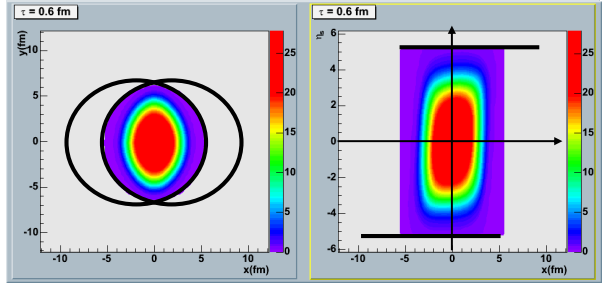}
\caption{Energy density distribution in a non-central H.I.C.
    within a CGC initial condition in the
    transverse plane (left panel) and in the reaction plane (right
    panel). The two horizontal thick black lines in the right panel
     are the Lorentz contracted nuclei.
     The color gradation in the right side of each panel
     indicates the energy density scale in unit of GeV/fm$^3$.}
\label{fig:slide65}
\end{figure}

\begin{figure}
\centering
\includegraphics[width=6cm]{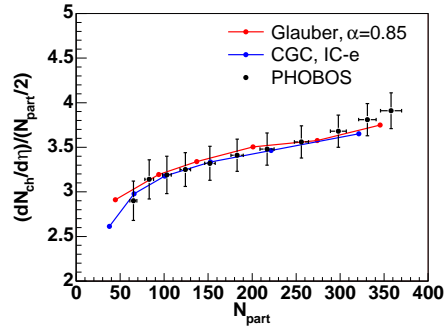}
\caption{Centrality dependence of multiplicity from
PHOBOS \cite{PHOBOS_dNdeta} are fitted by hydrodynamic calculations with two different initial
conditions \cite{Hirano:2005xf,Hirano:2004en}.}
\label{fig:slide66a}
\end{figure}

\begin{figure}
\centering
\includegraphics[width=6cm]{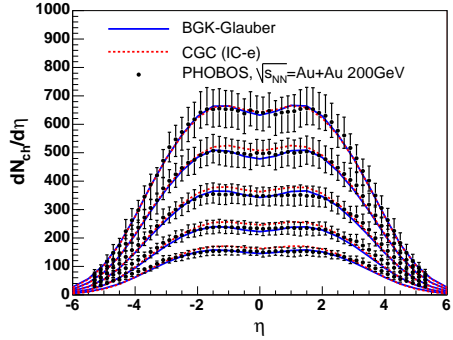}
\caption{Pseudorapidity dependence of multiplicity from
PHOBOS \cite{PHOBOS_dNdeta} are 
fitted by hydrodynamic calculations with two different initial
conditions \cite{Hirano:2005xf,Hirano:2004en}.}
\label{fig:slide66b}
\end{figure}

Initial conditions in hydrodynamic simulations
are so chosen as to reproduce the
centrality and rapidity dependences of multiplicity $dN_{\mathrm{ch}}/d\eta$.
Initial
conditions here mean energy density distribution $e(x,y,\eta_s)$ and flow velocity
$u^{\mu}(x,y,\eta_s)$ at the initial time $\tau_0$.
Again baryon density is neglected since, at midrapidity at RHIC, the net baryon
density is quite small.
The pressure distribution can be obtained from the energy density
distribution through the EoS.
Space-time rapidity $\eta_s$ independent
initial energy density distribution $e(x,y,\eta_s) = e(x,y)$ 
and Bjorken scaling solution $u^\mu_{\mathrm{Bj}}$
are assumed in (2+1)-dimensional hydrodynamic simulations.
In this case, one discuss the observables only at midrapidity.
At $\eta_s = 0$, one can parametrize \cite{Kolb:2001qz}
the initial entropy density based
on the Glauber model
\begin{equation}
s(x,y) = \frac{dS}{\tau_0 d\eta_s d^2x_\perp}  \propto 
\alpha n_{\rm part}(x,y; b) 
+ (1{-}\alpha)n_{\rm coll}(x,y; b)
\label{eq:entroini}
\end{equation}
The soft/hard fraction $\alpha$ is adjusted to reproduce the 
measured centrality dependence \cite{PHOBOS_Nch} of the charged hadron 
multiplicity at midrapidity.
By using the EoS, one can calculate the initial energy density distribution
from Eq.~(\ref{eq:entroini}).
For fully three-dimensional initial conditions, see Refs.~\cite{Hirano02,Hirano:2005xf,Hirano:2001eu}.
A novel initial condition is based on the color glass condensate (CGC)
picture \cite{Iancu:2003xm}.
One can calculate the local energy density of produced gluons within the CGC framework
\cite{Kharzeev:2001gp,Kharzeev:2002ei}
and utilize it as an initial condition of hydrodynamic simulations.
In Fig.~\ref{fig:slide65}, an
example of the CGC initial energy density distribution for a non-central H.I.C.
in a full (3+1)-dimensional hydrodynamic simulation  \cite{Hirano:2004en}
 is shown in the transverse plane (left) and in the reaction plane (right).
In the right side panel the horizontal axis corresponds to the impact parameter
direction,
and the vertical axis to the space-time rapidity $\eta_s$.
 Figures~\ref{fig:slide66a} and \ref{fig:slide66b} show charged particle
 multiplicity 
 from hydrodynamic simulations
 are compared with the PHOBOS data \cite{PHOBOS_Nch,PHOBOS_dNdeta}.
Figure~\ref{fig:slide66a} shows $dN_{\mathrm{ch}}/d\eta$ as a function of the number of
participants ($N_{\mathrm{part}}$) \cite{PHOBOS_Nch}.
This data is fitted by using two kinds of initial conditions;
from Glauber model calculations and from Color Glass Condensate (CGC) model
calculations \cite{Hirano:2005xf}.
Both models reproduce the centrality dependence of the data.
Figure~\ref{fig:slide66b} shows the rapidity distribution of 
$dN_{\mathrm{ch}}/d\eta$ for
each centrality \cite{PHOBOS_dNdeta}.
The fitting of multiplicity is the starting point of further analysis based on 
hydrodynamic simulations.

In the hydrodynamic models, various combinations of initial
conditions, EoS and decoupling prescriptions
are available to analyze the experimental data in H.I.C.
Of course, final results largely depend on modeling of each ingredient.
So it is quite important to constrain each model and its inherent parameters
through systematic analyses of the data toward a comprehensive understanding 
of the QGP.

\subsection{Application of the ideal hydrodynamic model to H.I.C. }

In this subsection we analyze H.I.C. at RHIC in terms of ideal hydrodynamic models
discussed in the previous subsection.

Before we start our main discussion on elliptic flow parameter $v_2$, we mention here that
the transverse momentum distributions for pions, kaons and protons are
also important since these reflect dominant transverse flow, namely radial flow.
Currently, among hydrodynamic models, 
yields and slopes of $p_{T}$ spectra
are reproduced in pure hydrodynamic calculations
with early chemical freezeout or in gradual freezeout approaches.
It should be noted here that simultaneous reproduction of the yields and the slopes is
important. Sometimes, one only compares the slope of the $p_T$ spectra by scaling
the yields ``by hand" within hydrodynamic approaches.
However, chemical composition of hadronic matter
does affect the transverse expansion \cite{Hirano02}.
Therefore, it does not make any senses
if one compares only the slopes by keeping chemical equilibrium of hadrons.

\begin{figure}
\centering
\includegraphics[width=6cm]{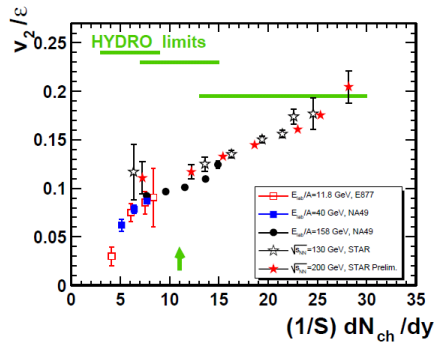}
\caption{$v_2/\varepsilon$ as a function of
transverse multiplicity density compiled by NA49 Collaboration \cite{Alt:2003ab}.}
\label{fig:slide70a}
\end{figure}

\begin{figure}
\centering
\includegraphics[width=6cm]{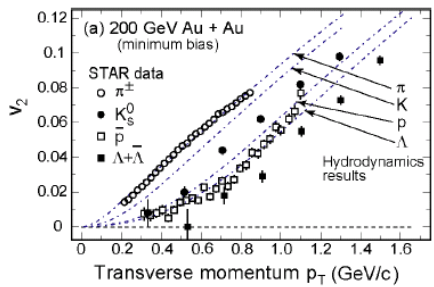}
\caption{Differential $v_2$ for pions, kaons, protons and lambdas \cite{Adams:2005dq}.}
\label{fig:slide70b}
\end{figure}

\begin{figure}
\centering
\includegraphics[width=6cm]{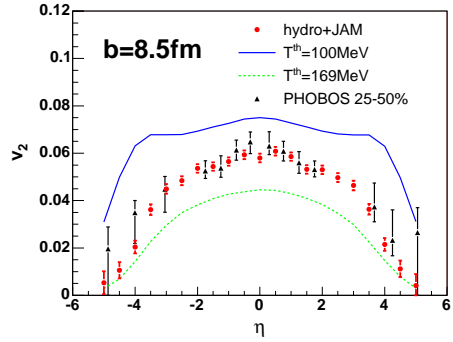}
\caption{Pseudorapidity dependence of $v_2$. PHOBOS data \cite{Back:2004mh} compared to different
model calculations \cite{Hirano:2005xf}.}
\label{fig:slide74}
\end{figure}

\begin{figure}
\centering
\includegraphics[width=6cm]{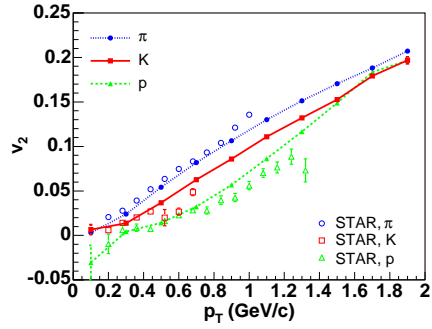}
\caption{Differential $v_2$. STAR data \cite{Adams:2005dq} compared to model calculations \cite{Hirano:2007ei}.}
\label{fig:slide75a}
\end{figure}

\begin{figure}
\centering
\includegraphics[width=1.0\linewidth]{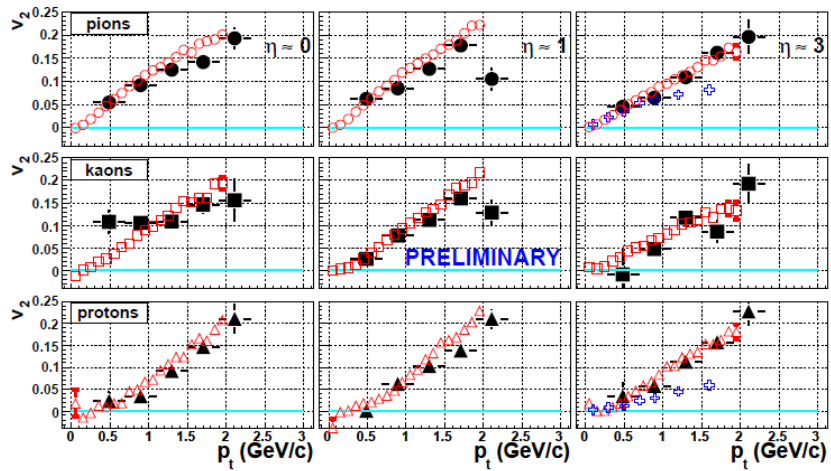}
\caption{Differential $v_2$ for pions, kaons and protons for $\eta=0$ (left), $\eta=1$ (middle) 
and $\eta=3$ (right) \cite{Sanders:2007th}.}
\label{fig:slide76}
\end{figure}

As discussed in Sec.~\ref{s:ellipticflow}, 
$v_2/\varepsilon$ can be interpreted as a response of the system 
to initial spatial eccentricity.
Figure~\ref{fig:slide70a} shows $v_2/\varepsilon$ as a function of 
the transverse multiplicity density $(1/S)dN_{\mathrm{ch}}/dy$
from AGS to RHIC energies.
Hydrodynamic results in Fig.~\ref{fig:slide60b}
are shown symbolically as horizontal lines.
The experimental data monotonically
increase with particle density,
while ideal hydrodynamic response is almost flat \cite{Kolb:2000sd}.
Ideal hydrodynamics is expected to generate the maximum
response among the transport models\footnote{
It should be emphasized again that
the hydrodynamic results above are obtained by
a particular combination of modeling, \textit{i.e.}, Glauber type initial conditions,
EOS Q with chemical equilibrium in the hadron phase and
sudden freezeout at fixed decoupling temperature.}.
The experimental data reach this limit for the first time at RHIC.
Figure~\ref{fig:slide70b} shows the differential elliptic flow,
$v_2$ as a function of transverse momentum for pions, kaons,
protons and lambdas. A mass ordering pattern is seen
in $v_2$ data, which was
predicted by ideal hydrodynamic calculations \cite{Huovinen:2001cy} 
\footnote{There is a caveat to interpret the agreement since
this particular hydrodynamic calculation does not reproduce
particle ratios due to a lack of early chemical freezeout.
The importance of hadronic viscosity and chemical freezeout
in hydrodynamic calculations
is recognized \cite{Hirano:2005wx}
after the announcement of the discovery of perfect fluid QGP \cite{BNL}.}.
The pseudorapidity dependence of $v_2$
observed by PHOBOS \cite{Back:2004mh} has a triangular shape as
is seen in Fig.~\ref{fig:slide74}. 
In the pure ideal hydrodynamic result, hydrodynamic equations are initialized by 
the Glauber model and are solved
all the way down
to $T^{\mathrm{dec}}=100$ MeV.
The pure hydrodynamic model
gives a comparable result with the data only at midrapidity.
However at forward and
backward rapidities, it overshoots the data significantly.
If we replace the
hadron fluid with a hadron gas utilizing a hadron cascade,
$v_2$ is significantly reduced in the forward and backward region.
In this hybrid model the hadrons have a finite mean free path,
which results in an effective shear viscosity in the hadron phase.
So dissipative hadronic ``corona" effects turn out to be important in understanding
the $v_2$ data.
The model also reproduces a mass ordering pattern of
$v_2$ for identified hadrons
as a function of $p_{T}$ near midrapidity in
Fig.~\ref{fig:slide75a}.
\begin{figure}
\centering
\includegraphics[width=6cm]{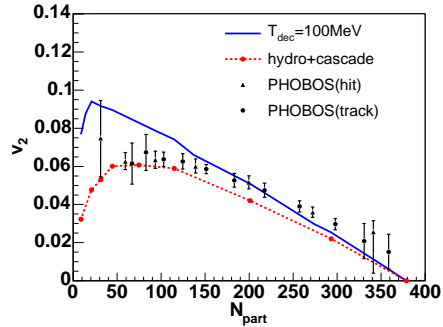}
\caption{$v_2$ as a function of centrality. PHOBOS data \cite{Back:2004mh} compared to different
model calculations \cite{Hirano:2005xf}.}
\label{fig:slide75b}
\end{figure}
Figure~\ref{fig:slide75b} shows the centrality dependence of $v_2$.
The solid line is the result from ideal hydrodynamic calculations, 
while the dotted line from the
hybrid model.
It is clear that for peripheral collisions, where the multiplicity is small,
the hadronic viscosity plays an important role.
One may notice that the result from the hybrid model systematically and slightly smaller
than the data. However, there could exist the effect of initial eccentricity
fluctuations which is absent in this hydrodynamic calculations.
The deviation between the results and the data can be interpreted quantitatively
by this effect.
Figure~\ref{fig:slide76} shows $v_2(p_T)$ 
for pions,
kaons and protons
in 10-50\% centrality
at $\eta=0$ (left), $\eta=1$ (middle) and $\eta=3$ (right)
observed by BRAHMS \cite{Sanders:2007th}.
Also here the hybrid model reproduces the $p_{T}$ slope of these differential
elliptic flow parameters.

\begin{figure}
\centering
\includegraphics[width=6cm]{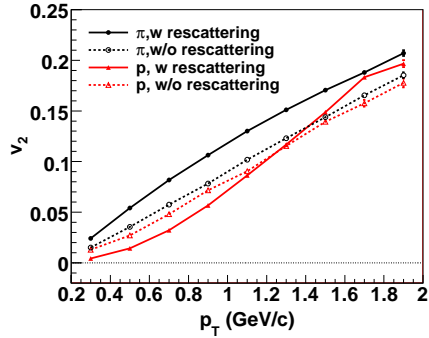}
\caption{Differential $v_2$ with and without hadronic rescattering \cite{Hirano:2007ei}.}
\label{fig:slide77b}
\end{figure}

We would like to point out here that the mass ordering, clearly
visible in Fig.~\ref{fig:slide75a}, is there in the final
result. If one would look at the result just after the QGP phase
transition, the difference between the pions and protons would be
quite small.
So it turns out that the splitting patterns are caused by hadronic rescattering.
This is illustrated in Fig.~\ref{fig:slide77b}.
One can conclude that the large magnitude of the 
integrated $v_2$ and the strong mass ordering of the differential 
$v_{2}(p_{T})$ observed at RHIC result from a subtle interplay between 
perfect fluid dynamics of the early QGP stage and dissipative dynamics 
of the late hadronic stage: The large magnitude of $v_2$ is due to 
the large overall momentum anisotropy, generated predominantly in the 
early QGP stage, whereas the strong mass splitting behavior
at low $p_T$ reflects the redistribution of this momentum 
anisotropy among the different hadron species, driven by the 
continuing radial acceleration and cooling of the matter during the 
hadronic rescattering phase.

\begin{figure}
\centering
\includegraphics[width=6cm]{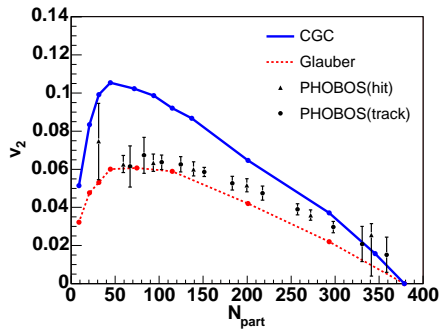}
\caption{$v_2$ as a function of centrality. PHOBOS data \cite{Back:2004mh}
are compared to hydrodynamic results with two different sets of initial conditions \cite{Hirano:2005xf}.}
\label{fig:slide79}
\end{figure}

\begin{figure}
  \centerline{
    \begin{tabular}{cp{0.25cm}c}
      \subfigure[Energy density distribution]{
        \includegraphics[width=\subfigsize]{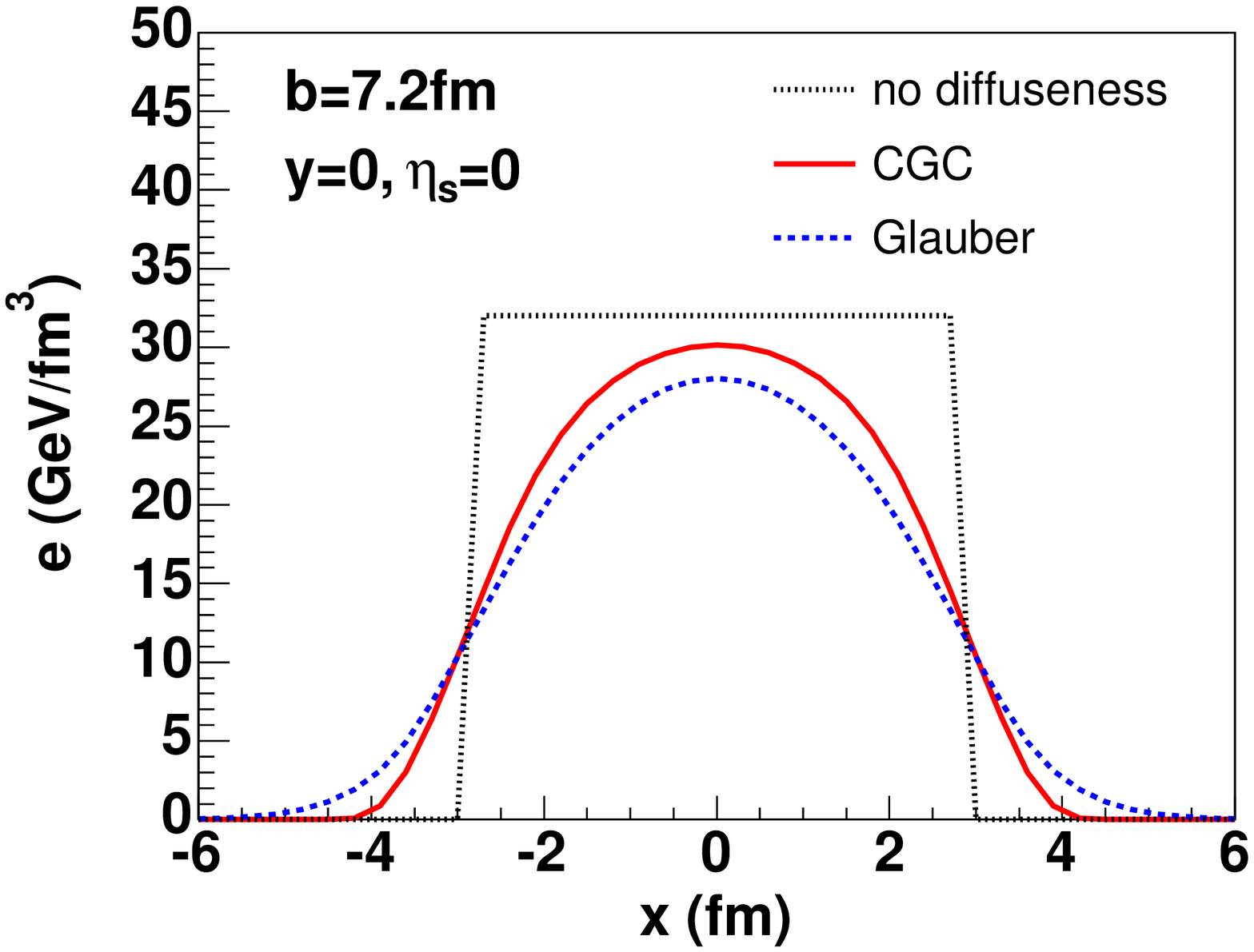}
        \label{fig:slide80b}
      } & &
      \subfigure[Initial eccentricity]{
        \includegraphics[width=\subfigsize]{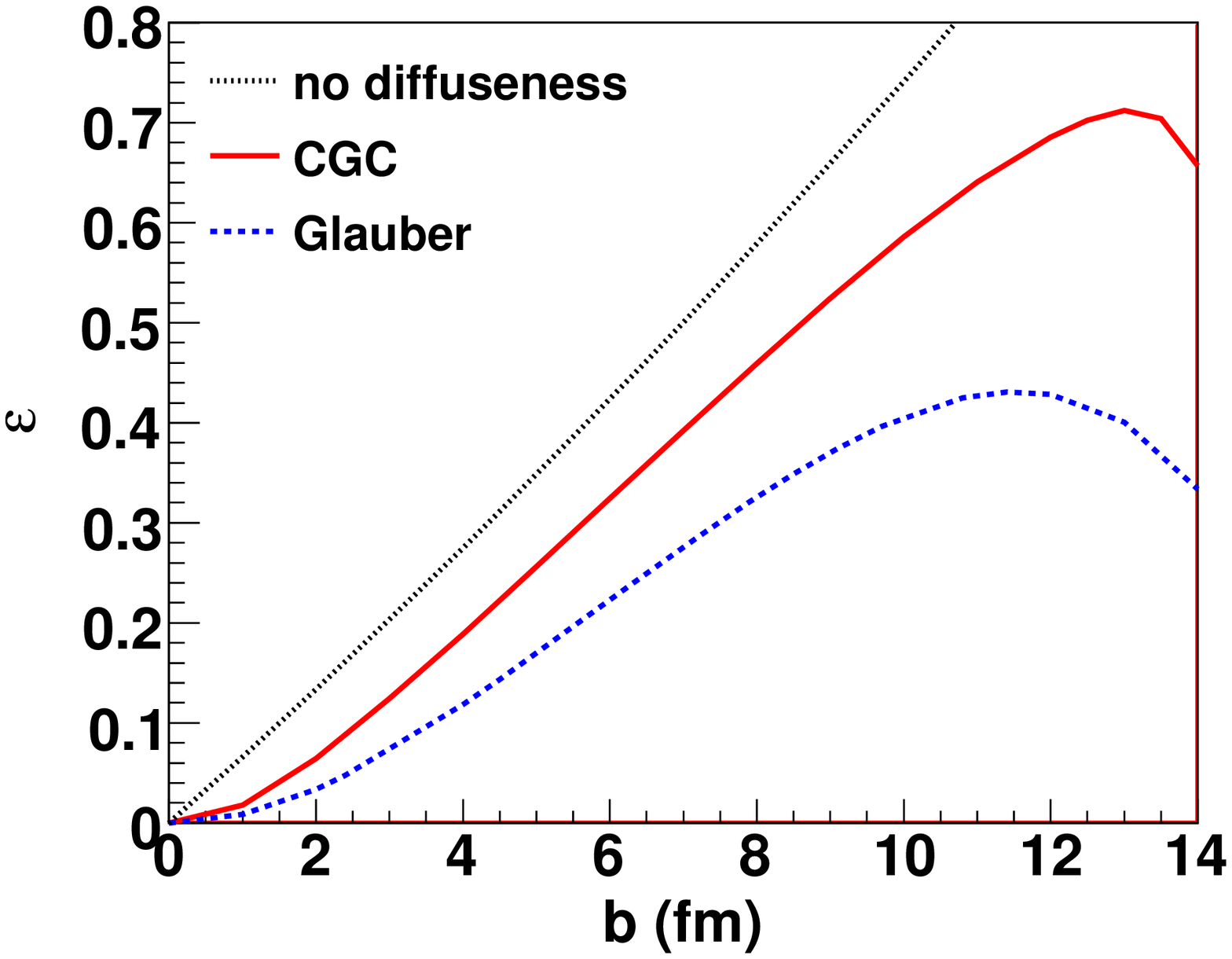}
        \label{fig:slide80c}
      }
    \end{tabular}
  }
  \caption{The difference eccentricity between Glauber and CGC initial conditions.}
  \label{fig:slide80}
\end{figure}

We have seen so far that the hydrodynamic model which includes
Glauber type initial conditions followed by a perfect fluid QGP
and a dissipative hadronic gas evolution is the most successful
combination for describing the RHIC data.
We now go to the discussion on the initialization dependence of $v_{2}$.
Two types of initial conditions, namely the Glauber type initial conditions
and the CGC initial conditions, are discussed in the previous subsection.
$v_2$ as a function of centrality
is shown again for these two initial conditions in Fig.~\ref{fig:slide79}.
In the case of the Glauber initial conditions we can conclude early thermalization
and the discovery for the perfect fluid QGP.
In the case of the CGC initial conditions,  we cannot, however,
claim the discovery since the model initialized
by CGC overshoots the data in almost the whole range.
Since the hydrodynamic model calculations depend on the initial conditions,
it is very important to understand them before making final conclusions.
In the case of CGC initial conditions viscosity might be needed 
even in the QGP phase to
get the model
down to the data points.
The effect of viscosity could therefore be quite important.
The high $v_2$ values from the CGC initial conditions are 
traced back to the initial eccentricity.
In Fig.~\ref{fig:slide80b} the energy density distribution
in the impact parameter direction is plotted for
different conditions.
If the energy density profile has a sharp edge (no diffuseness),
an integral in Eq.~(\ref{eq:stdecc}) is relatively weighted in the edge region and, consequently,
eccentricity becomes maximum at a given impact parameter.
If one compare the energy density profile of the CGC with the one of the Glauber model,
one see the CGC profile has a sharper edge than the Glauber model does.
The resultant eccentricity as a function of impact parameter is
shown in Fig.~\ref{fig:slide80c}.
Eccentricity from the CGC is about 20-30\% larger than that from the Glauber model.
This is the reason why hydro + hadronic cascade approach which even includes
hadronic viscosity overshoots the $v_2$ data~\footnote{See also recent more realistic
calculations of eccentricity within a CGC framework \cite{Drescher:2006ca}.}.

\subsection{Summary}

Hydrodynamics is a framework to describe
the space-time evolution of matter
under local equilibrium. 
It is applied to the intermediate stage in H.I.C.
to extract the transport properties of 
the QGP from RHIC data.
Hydrodynamic modeling includes initial conditions, EoS and
decoupling prescriptions.
Final results certainly depend on combination of each modeling.
So much attention should be paid to
these ingredients before drawing robust conclusions from hydrodynamic
analyses.
Elliptic flow has played a mayor role in understanding the transport properties of
the QGP.
Glauber initial conditions, ideal hydrodynamics in the QGP phase and dissipative gas
for the hadron phase
are three pillars for agreement between the model and elliptic flow data.
Whereas, if CGC initial conditions are employed,
the initial eccentricity gets increased by 20-30\%.
If the nature chooses this kind of initial condition,
viscosity might be needed even in the QGP phase.

\section*{Acknowledgments}
One of the authors (T.H.) is much indebted to
M.~Gyulassy, T.~Hatsuda, U.~Heinz, T.~Kunihiro, S.~Muroya, M.~Natsuume and Y.~Nara
 for fruitful discussion.
The work of T.H. was partly supported by Grant-in-Aid for Scientific
Research No.~19740130.
One of the authors (A.B.) would like to thank Despoina Evangelakou for technical support during the writing of the lecture note.

%
%

%
%



\printindex
\end{document}